\documentclass[aps,amsmath,twocolumn,amssymb,floatfixng,showpacs,
superscriptaddress,footinbib]{revtex4-2}
\usepackage{bm}
\usepackage{graphicx}
\usepackage{graphics}
\usepackage{mathtools}
\usepackage{amsmath}
\usepackage{newtxtext} 
\usepackage{epstopdf}
\usepackage{hyperref}
\hypersetup{
    colorlinks,%
    citecolor=blue,%
    linkcolor=blue,%
    urlcolor=blue
}
\usepackage{tikz}\usetikzlibrary{petri}
\usepackage{soul}

\DeclareMathOperator{\Tr}{Tr}

\begin{document}

\newcommand{\be}   {\begin{equation}}
\newcommand{\ee}   {\end{equation}}
\newcommand{\ba}   {\begin{eqnarray}}
\newcommand{\ea}   {\end{eqnarray}}
\newcommand{\ve}  {\varepsilon}
\newcommand{\bbs}[1]{{\boldsymbol{{#1}}}} 

\title{ Signatures of multiorbital density waves in the electronic Raman response of bilayer nickelates }

\author{Lauro B. Braz}
\affiliation{
Instituto de F\'{\i}sica, Universidade de S\~ao Paulo, Rua do Mat\~ao 1371, S\~ao Paulo, S\~ao Paulo 05508-090, Brazil
}
\affiliation{
Theoretische Physik III, Ruhr-Universit\"at Bochum, D-44780, Bochum, Germany
}
\author{Ge He}
\affiliation{School of Mechanical Engineering, Beijing Institute of Technology, Beijing 100081, China}
\affiliation{Beijing Key Laboratory of Quantum Matter State Control and Ultra-Precision Measurement Technology, Beijing Institute of Technology, Beijing 100081, China}
\author{Luis G.~G.~V. Dias da Silva}
\affiliation{
Instituto de F\'{\i}sica, Universidade de S\~ao Paulo, Rua do Mat\~ao 1371, S\~ao Paulo, S\~ao Paulo 05508-090, Brazil
}
\author{Ilya M. Eremin}
\affiliation{
Theoretische Physik III, Ruhr-Universit\"at Bochum, D-44780, Bochum, Germany
}

\date{ \today }

\begin{abstract}

Recent Raman scattering experiments on bilayer nickelates have revealed pronounced spectral features in the spin-density-wave state, whose microscopic origin remains controversial, with interpretations ranging from two-magnon excitations to electronic transitions across SDW-induced gaps. Here, we demonstrate that the multiorbital electronic structure of low-pressure bilayer nickelates naturally generates distinct orbital-selective Raman signatures in different symmetry channels. Using a two-orbital bilayer model incorporating the $d_{3z^2-r^2}$ and $d_{x^2-y^2}$ orbitals in the orthorhombic structure, we analyze the electronic Raman response  of the multiorbital spin-spinless double stripe SDW state and show that the $B_{2g}$ response exhibits a peak at the energy scale $(M_{z^2}+M_{x^2})/2$, associated with SDW gap openings in weakly hybridized bands. In contrast, the dominant peak in the $B_{1g}$ channel emerges at the multiorbital energy scale $(M_{z^2}^2+M_{x^2}^2)^{1/2}$, reflecting the strongly hybridized character of the relevant electronic states. These distinct energy scales provide a unified interpretation of several Raman signatures observed in La$_3$Ni$_2$O$_7$ and offer a means of disentangling the orbital contributions to the SDW order parameters. Our results highlight that in contrast to the superconducting state, Raman peak positions in the SDW state cannot generally be identified with a single SDW gap and establish symmetry-resolved electronic Raman scattering as a sensitive probe of multiorbital density-wave order.
\end{abstract}

\maketitle

\section{Introduction}
Electronic Raman spectroscopy has played a pivotal role in elucidating the intertwined electronic and magnetic phenomena underlying unconventional superconductivity, providing insights that extend well beyond the superconducting state itself. In cuprate superconductors, Raman scattering revealed the persistence of short-range antiferromagnetic correlations well into the doped regime, including the superconducting phase, highlighting the possible importance of magnetic fluctuations for the pairing mechanism \cite{Reznik1993PlanePolarizedRaman,Blumberg1994TwoMagnonRaman,Rubhausen1999DopingDependence,Sugai2003CarrierDensityRaman,Devereaux2007InelasticLightScattering}. Moreover, the pronounced polarization dependence of the Raman response provided some of the earliest spectroscopic evidence for an anisotropic superconducting gap \cite{Cooper1988RamanScattering,Cooper1988GapAnisotropy,Hackl1988LightScattering,Yamanaka1988SuperconductingGapExcitations,Slakey1990RamanSuperconductivity}, subsequently established to have predominantly $d_{x^2-y^2}$ symmetry \cite{Scalapino1995CaseDx2Y2Pairing}. In iron-based superconductors, symmetry-resolved Raman spectroscopy has likewise emerged as a powerful probe of nematic fluctuations, revealing a strongly enhanced response in the $B_{1g}$ symmetry channel even in the absence of long-range nematic order \cite{Gallais2013IncipientChargeNematicity,Kretzschmar2016CriticalSpinFluctuations,Gallais2016ChargeNematicity,Hinojosa2016RamanResonance,Karahasanovic2015ChargeDrivenNematicity,Khodas2015RamanNematicCorrelations,Kontani2014LinearResponseTheory,Thorsmolle2016CriticalQuadrupoleFluctuations,Mancini2017IronPnictideSuperconductors}. These examples demonstrate how the symmetry selectivity of Raman scattering provides access to distinct electronic and collective excitations, making it particularly valuable for investigating materials in which superconductivity competes or coexists with other ordered states.

The symmetry-selective sensitivity of Raman spectroscopy to electronic and collective excitations is particularly relevant to layered nickelates, where superconductivity emerges in close proximity to competing density-wave orders. At ambient and low pressures, both bilayer and trilayer nickelates exhibit spin-density-wave (SDW) transitions, accompanied by charge-density-wave (CDW) order in the trilayer compounds \cite{Puphal2026NickelateReview}. In trilayer nickelates, Raman measurements reveal a pronounced depletion of the electronic continuum upon entering the density-wave state \cite{Gim2025OrbitalSelectiveQuasiparticleDepletion,Sundaramurthy2025ComparativeRaman,Suthar2026MultiorbitalDensityWave}, together with the appearance of additional phonon modes associated with the accompanying lattice reconstruction \cite{Suthar2026MultiorbitalDensityWave,Deswal2025DynamicsElectronCorrelation}. In the bilayer compound La$_3$Ni$_2$O$_7$, Raman experiments have revealed pronounced and strongly symmetry-dependent spectral features below the SDW transition. In particular, the $B_{1g}$ response exhibits a stronger intensity and a dominant peak at higher energy than the corresponding $B_{2g}$ feature \cite{He2026AnisotropicNickelate,Gim2026SpectroscopicNickelate}. The two channels also display distinct line shapes, with a relatively broad, approximately Gaussian $B_{1g}$ peak and a sharper $B_{2g}$ feature accompanied by a higher-energy shoulder or tail \cite{He2026AnisotropicNickelate,Gim2026SpectroscopicNickelate,Zhou2026UnconventionalPressureEvolution}. Moreover, pressure-dependent measurements reveal a pronounced suppression of the $B_{2g}$ response, while the $B_{1g}$ signal remains comparatively robust \cite{Zhou2026UnconventionalPressureEvolution}. Despite these consistent experimental observations, their microscopic interpretation remains controversial. He et al. \cite{He2026AnisotropicNickelate} attribute the Raman features to electronic particle--hole excitations across SDW-induced gaps in the reconstructed band structure, whereas Gim et al. \cite{Gim2026SpectroscopicNickelate} interpret the prominent spectral peaks primarily in terms of two-magnon excitations associated with competing magnetic exchange interactions. This ambiguity is particularly significant in a multiorbital system, where the Raman response can probe different regions of the reconstructed electronic structure depending on the polarization geometry. Furthermore, strong phonon resonances can obscure parts of the electronic continuum, complicating the identification of the underlying excitations. Consequently, a microscopic understanding of the symmetry-dependent Raman response in the SDW state is essential for disentangling electronic reconstruction effects from collective magnetic excitations.

An additional complication arises from the sensitivity of the electronic structure of bilayer nickelates to structural distortions and pressure. Phonon Raman spectroscopy has been employed to investigate structural transitions involving the tetragonal $I4/mmm$ and orthorhombic $Amam$ or $Am2m$ structures, with pressure stabilizing the high-pressure orthorhombic $Fmmm$ phase \cite{Zhang2025IdentifyingStructureLa3Ni2O7,Puphal2026NickelateReview}. While the $I4/mmm$ and $Fmmm$ structures exhibit approximately straight interlayer Ni--O--Ni bonds, the low-pressure $Amam$ and $Am2m$ structures feature tilted NiO$_6$ octahedra. These distortions modify the orbital hybridization and electronic band structure, potentially influencing both the stability of density-wave order and its Raman signatures. An additional aspect of this problem is the multiorbital nature of the electronic structure in bilayer nickelates. The low-pressure orthorhombic structure, characterized by tilted NiO$_6$ octahedra, modifies the hybridization between the Ni 3$d_{3z^2-r^2}$ and 3$d_{x^2-y^2}$ orbitals and as a result influences the reconstruction of the electronic bands in the SDW state \cite{Zhang2025IdentifyingStructureLa3Ni2O7,Puphal2026NickelateReview}. Since the electronic Raman response depends on the orbital character and momentum dependence of the Raman vertices, different symmetry channels can selectively probe distinct parts of the reconstructed band structure. This raises the question of whether the experimentally observed differences between the $B_{1g}$ and $B_{2g}$ responses can be understood within a multiorbital description of electronic excitations in the SDW state.

In this work, we follow this route and investigate the electronic Raman response in the SDW state of low-pressure bilayer nickelates, aiming to clarify the microscopic origin of the experimentally observed symmetry-dependent spectral features. Starting from a two-orbital bilayer model incorporating the Ni $d_{3z^2-r^2}$ and $d_{x^2-y^2}$ orbitals, we examine how orbital-dependent SDW order parameters and interorbital hybridization determine the Raman response in the $B_{1g}$ and $B_{2g}$ symmetry channels. We demonstrate that the electronic reconstruction induced by SDW order naturally generates two distinct characteristic energy scales. In the $B_{2g}$ channel, the dominant spectral feature originates predominantly from weakly hybridized bands and appears at an energy scale approximately given by the individual orbital-resolved SDW gaps $(M_{z^2}+M_{x^2})/2$. In contrast, the $B_{1g}$ response is dominated by strongly hybridized multiorbital states, giving rise to a peak at the energy scale $(M_{z^2}^2+M_{x^2}^2)^{1/2}$. These results provide a microscopic electronic interpretation of the distinct Raman features observed in La$_3$Ni$_2$O$_7$. We show that the characteristic energies, seen in Raman spectroscopy cannot, in general, be directly identified with individual SDW gaps. While our analysis does not exclude additional contributions from collective magnetic excitations, it establishes that the experimentally observed separation of Raman energy scales can arise naturally from the multiorbital electronic structure of the SDW state. More generally, our findings highlight the potential of symmetry-resolved Raman spectroscopy to disentangle orbital-dependent density-wave order parameters in multiorbital correlated materials.

\section{Theoretical model of the double stripe SDW state}
The crystal structure of bilayer nickelates at low pressure has been the subject of considerable recent discussion. Early experiments identified an orthorhombic $Amam$ structure \cite{sunSignaturesSuperconductivity802023}, in which rotations and tilts of the NiO$_6$ octahedra enlarge the crystallographic unit cell to four Ni sites per bilayer. More recently, high-resolution synchrotron X-ray diffraction studies have suggested that the structural data are better described by the lower-symmetry $Am2m$ space group, which lacks a mirror symmetry of the $Amam$ structure \cite{Misawa2026,Reiss2026,Liu2026DensityWavePhases,Sasaki2026StructuralPhaseDiagram,Yashima2026LargeTinySpinStripe}. This symmetry reduction is accompanied by an nonequivalent in-plane Ni sites and has been associated with charge disproportionation persisting to temperatures as high as 544 K \cite{Zhang2025IdentifyingStructureLa3Ni2O7,Michon2026MetallicCrossover}. Thus, to describe the electronic structure in the low-pressure regime, we start from an eight-band tight-binding model constructed for the orthorhombic $Amam$ phase \cite{Wang2024ElectronicMagneticNickelate,Braz2026DensityWaves}. The model includes the two Ni $e_g$ orbitals, $d_{x^2-y^2}$ and $d_{3z^2-r^2}$, on each of the two inequivalent in-plane Ni sites, Ni$_1$ and Ni$_2$, and in both layers of the bilayer, denoted by $A$ and $B$. 
The structural differences between the $Amam$ and $Am2m$ phases are relatively small, whereas the associated charge disproportionation introduces an additional electronic energy scale. As a first approximation, we therefore retain the hopping parameters of the $Amam$ model and incorporate the leading effect of the reduced symmetry through a staggered on-site potential acting on the two inequivalent in-plane Ni sites. The resulting Hamiltonians are
\begin{align}
H_{Amam}
&=
\sum_{\boldsymbol{k}}
\sum_{\mu\eta}
\sum_{\sigma}
\xi^{\mu\eta}_{\boldsymbol{k}}
\, c^{\dagger}_{\boldsymbol{k}\mu\sigma}
c_{\boldsymbol{k}\eta\sigma}, \label{eq:H0_sublattice} \\
H_{Am2m} &= H_{Amam} + \frac{1}{2}\Delta_{\text{CO}}\tau^z\otimes\hat{1}_{\text{bilayer}},
\label{eq:am2m}
\end{align}
where $\tau^z$ is a Pauli matrix in the in-plane site space $\{\text{Ni}_1,\text{Ni}_2\}$, $\hat{1}_{\text{bi}}$ denotes the identity matrix in the bilayer two-orbital space $\{Ad_{x^2-y^2},Ad_{3z^2-r^2},Bd_{x^2-y^2},Bd_{3z^2-r^2}\}$, and $\Delta_{\text{CO}}$ is the checkerboard charge-order energy scale. Within this model, the checkerboard CDW becomes the ``normal state'' of the system.
Here, $\xi^{\mu\eta}_{\boldsymbol{k}}=\epsilon^{\mu\eta}_{\boldsymbol{k}}-\mu\delta_{\mu\eta}$, in which we define the chemical potential $\mu$ and the tight-binding matrix $\epsilon^{\mu\eta}_{\boldsymbol{k}}$ that includes hoppings up to third-nearest neighbors and a tuning parameter $\Delta_{xz}=\epsilon_{d_{x^2}}-\epsilon_{d_{z^2}}$ defined as the difference between on-site energies {(crystal-field splitting)} for $d_{x^2-y^2}$ and $d_{3z^2-r^2}$ orbitals.
DFT calculations based on experimentally-fitted crystal structures have shown that the ``crystal-field'' splitting parameter is the leading effect at low pressures \cite{Braz2026DensityWaves}.
Therefore, $\Delta_{xz}$ is hereby used to tune pressure.
$c_{\boldsymbol{k}\mu\sigma}$ annihilates an electron with momentum $\boldsymbol{k}$ and spin $\sigma$ in orbital $\mu$ indices.
This model considers the two-layer structure ($A$ and $B$), the two nonequivalent in-plane sites Ni$_1$ and Ni$_2$, and the two Ni orbitals $d_{x^2-y^2}$ and $d_{3z^2-r^2}$, totalizing eight bands.

Interactions are included using the field theory developed in Appendix~\ref{appendix:theory}.
A Hubbard-Kanamori Hamiltonian including onsite Hubbard $U$, Hund's coupling $J_H$, and interorbital $U'=U-2J_H$ is treated in a mean-field level, developed in Appendix \ref{appendix:derivation} and the unrestricted Hartree-Fock problem is solved by minimizing the free energy functional obtained from searching for saddle-point solutions of the multiorbital action written in terms of the magnetic $M_\mu$ and charge $\phi_\mu$ Hubbard-Stratonovich fields, which have physical interpretation in terms of the spin $\langle S_\mu\rangle$ and charge $\langle n_\mu\rangle$ densities,
\begin{align}
\phi_{\mu}
&=
\frac{1}{2}
\sum_{\eta}
(U_c)^{\mu\mu}_{\eta\eta}
\langle n_{\eta} \rangle,
\\
{M}_{\mu}
&=
\frac{1}{2}
\sum_{\eta}
(U_s)^{\mu\mu}_{\eta\eta}
\langle {S}_{\eta} \rangle,
\end{align}
where $n_{\eta}=
\sum_{\boldsymbol{k}\sigma}
c^{\dagger}_{\boldsymbol{k}\eta\sigma}
c_{\boldsymbol{k}\eta\sigma}$ is defined at $\boldsymbol{q}=\boldsymbol 0$ and ${S}_{\eta}=
\sum_{\boldsymbol{k}\sigma\sigma'}
c^{\dagger}_{(\boldsymbol{k}+\boldsymbol{q})\eta\sigma}
{\sigma}_{\sigma\sigma'}^z
c_{\boldsymbol{k}\eta\sigma'}$ at $\boldsymbol{q}=\boldsymbol{Q}$ the magnetic ordering vector.
${\sigma^z}$ denotes the spin-space Pauli matrix.
$\hat{U}_c$ and $\hat{U}_s$ are the charge and spin multiorbital interaction matrices, as defined in \cite{Braz2026DensityWaves}.
We have checked that the orthorhombic structure induces a single-$\bbs{Q}_Y=(0,\pi)$ order.
The mean-field Hamiltonian matrix reads \cite{Braz2026DensityWaves}
\begin{equation}
\hat h^{\sigma}(\boldsymbol k)=
\begin{pmatrix}
\hat\xi_{\boldsymbol k}+\hat\phi & -\sigma \hat{M} \\
-\sigma\hat M & \hat\xi_{\boldsymbol k+\boldsymbol Q}+\hat\phi
\end{pmatrix},\label{eq:HMF}
\end{equation}
which is diagonal in spin space. The spinful Hamiltonian becomes $\hat h=\text{diag}(\hat h^{\uparrow},\hat h^{\downarrow})$.
This Hamiltonian matrix is diagonalized as $\hat h(\boldsymbol k)|\boldsymbol{k}\nu\rangle=\epsilon_{\boldsymbol k\nu}|\boldsymbol{k}\nu\rangle$.

\begin{figure}[t]
\begin{center}
\includegraphics[width=1.0\columnwidth]{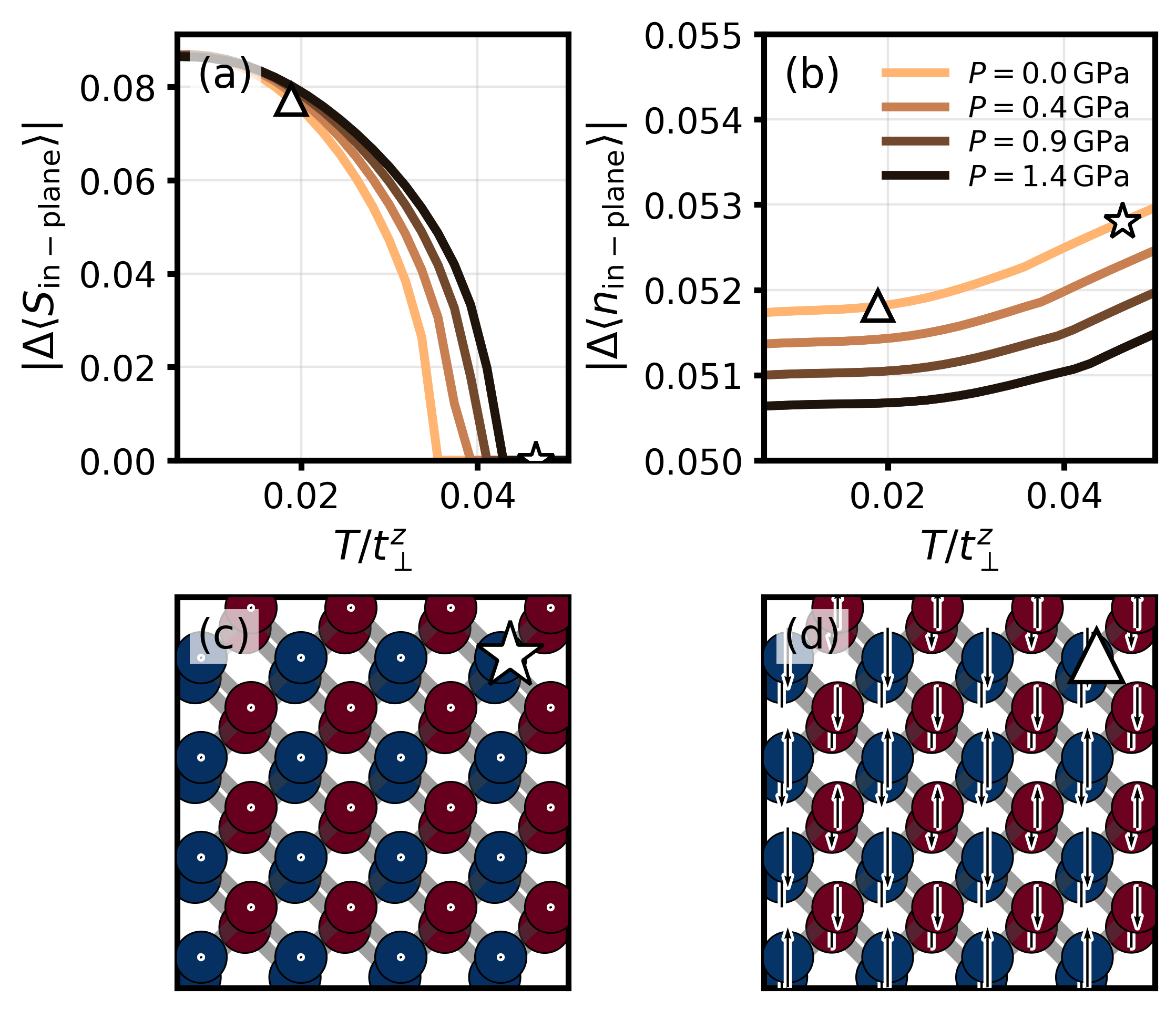}
\caption{ Calculated temperature and pressure dependence of the spin- and charge-density modulations in the orthorhombic $Am2m$ bilayer nickelate model. (a) Difference in the in-plane spin densities, $|\Delta\langle S_{\text{in-plane}}\rangle|$, between the two inequivalent Ni sites, showing the suppression of SDW order at the transition temperature $T_{\text{SDW}}(P)$; (b) shows corresponding difference in the local electron densities, $|\Delta\langle n_{\text{in-plane}}\rangle|$, which remains finite above the SDW transition due to the underlying charge disproportionation. 
Different colors indicate the applied pressures, as specified in the legend, with the charge-order potential fixed at $\Delta_{\text{CO}}/U=0.06$. Temperature is normalized by the interlayer hopping integral $t_\perp^z$.
(c,d) Schematic real-space configurations of the charge and spin densities corresponding to the star and triangle markers in panels (a,b), respectively. Blue and red circles represent positive and negative deviations of the local charge density from its spatial average, while arrows indicate the orientation of the local spin polarization. Panel (c) illustrates the charge-disproportionated state without long-range SDW order, whereas panel (d) shows the coexistence of charge disproportionation and SDW order.
}
\label{fig:op}
\end{center}
\end{figure}

Below $T_{\text{SDW}}$, double stripe spin-large spin-low magnetic order with ordering vector $\boldsymbol{Q}_Y=(0,\pi)$ develops on top of the pre-existing charge-disproportionated state. Owing to the inequivalence of the two in-plane Ni sites in the $Am2m$ structure, the onset of magnetic order is accompanied by a modulation of the local spin polarization within the four-Ni unit cell. Figure~\ref{fig:op}(a) shows the temperature dependence of the difference between the local spin densities on the two inequivalent Ni sites for several pressures. This difference vanishes at $T_{\text{SDW}}$ and increases continuously upon cooling, reflecting the development of the spin-modulated (spin-large spin-low) double stripe state. Increasing pressure reduces the magnitude of the spin-density modulation at a given reduced temperature, indicating a progressive weakening of the inequivalence between the local magnetic moments. The spin modulation is closely connected to the underlying charge disproportionation. As shown in Fig.~\ref{fig:op}(b), the difference in the local electron densities remains finite above $T_{\text{SDW}}$, reflecting the charge order already present in the high-temperature $Am2m$ phase. Upon entering the SDW state, the charge disproportionation is further enhanced, with the strongest effect occurring at ambient pressure. This enhancement becomes progressively weaker with increasing pressure. Importantly, the charge-order potential $\Delta_{\text{CO}}/U=0.06$ is kept fixed throughout these calculations. The pressure dependence of the charge-density difference therefore arises from the redistribution of electronic occupations associated with changes in the underlying band structure, rather than from an externally imposed reduction of the charge-order potential. The additional enhancement below $T_{\text{SDW}}$ demonstrates a self-consistent feedback between the spin and charge sectors.

The resulting real-space configurations are illustrated in Figs.~\ref{fig:op}(c) and (d). Above $T_{\text{SDW}}$, the system exhibits charge disproportionation between the nonequivalent Ni sites without long-range magnetic order [Fig.~\ref{fig:op}(c)]. Below the transition, double stripe magnetic order develops with unequal local moment amplitudes on the two charge-inequivalent sites [Fig.~\ref{fig:op}(d)]. Within the present solution, the charge-rich sites (blue circles) support larger local spin polarizations, whereas the charge-poor sites (red circles) exhibit reduced moments. This correlation reflects the itinerant and multiorbital character of magnetism in bilayer nickelates: the electronic occupations and local spin polarizations are determined self-consistently by the same reconstructed electronic states. Consequently, the pre-existing charge disproportionation provides a natural mechanism for generating a spin-amplitude modulation within the stripe-ordered phase. We note, that this scenario is consistent with experimental indications of inequivalent local magnetic environments. In particular, $\mu$SR measurements reveal distinct internal-field components whose pressure evolution suggests a redistribution of local magnetic fields, even as the SDW transition temperature increases \cite{Khasanov_PressureEnhancedSplittingDensityWaveTransitions_2025,Khasanov2026IsotopeNickelate}. Complementary $^{139}$La nuclear quadrupole resonance measurements provide further evidence for an inhomogeneous magnetic state compatible with spin-modulated stripe order \cite{Yashima2026LargeTinySpinStripe}. Our calculations offer a microscopic interpretation of these observations in terms of the coupling between pre-existing charge disproportionation and itinerant stripe magnetism.

We further examined the role of the pre-existing charge order in Appendix~\ref{appendix:co}, where we compare the results with calculations performed for $\Delta_{\text{CO}}/U=0$. In the absence of an imposed charge-order potential, the model yields either an unmodulated double-stripe magnetic solution or a charge modulation that develops only within the SDW phase \cite{Braz2026DensityWaves}. The latter scenario would associate the appearance of inequivalent local magnetic environments directly with the magnetic transition, rather than with a charge-disproportionated state already established at higher temperatures. The experimentally observed evolution of the local magnetic fields therefore favors a description in which the $Am2m$ charge disproportionation precedes the SDW transition and acts as a symmetry-breaking background for the subsequent spin-modulated stripe order.

\begin{figure}[t]
\begin{center}
\includegraphics[width=1.0\columnwidth]{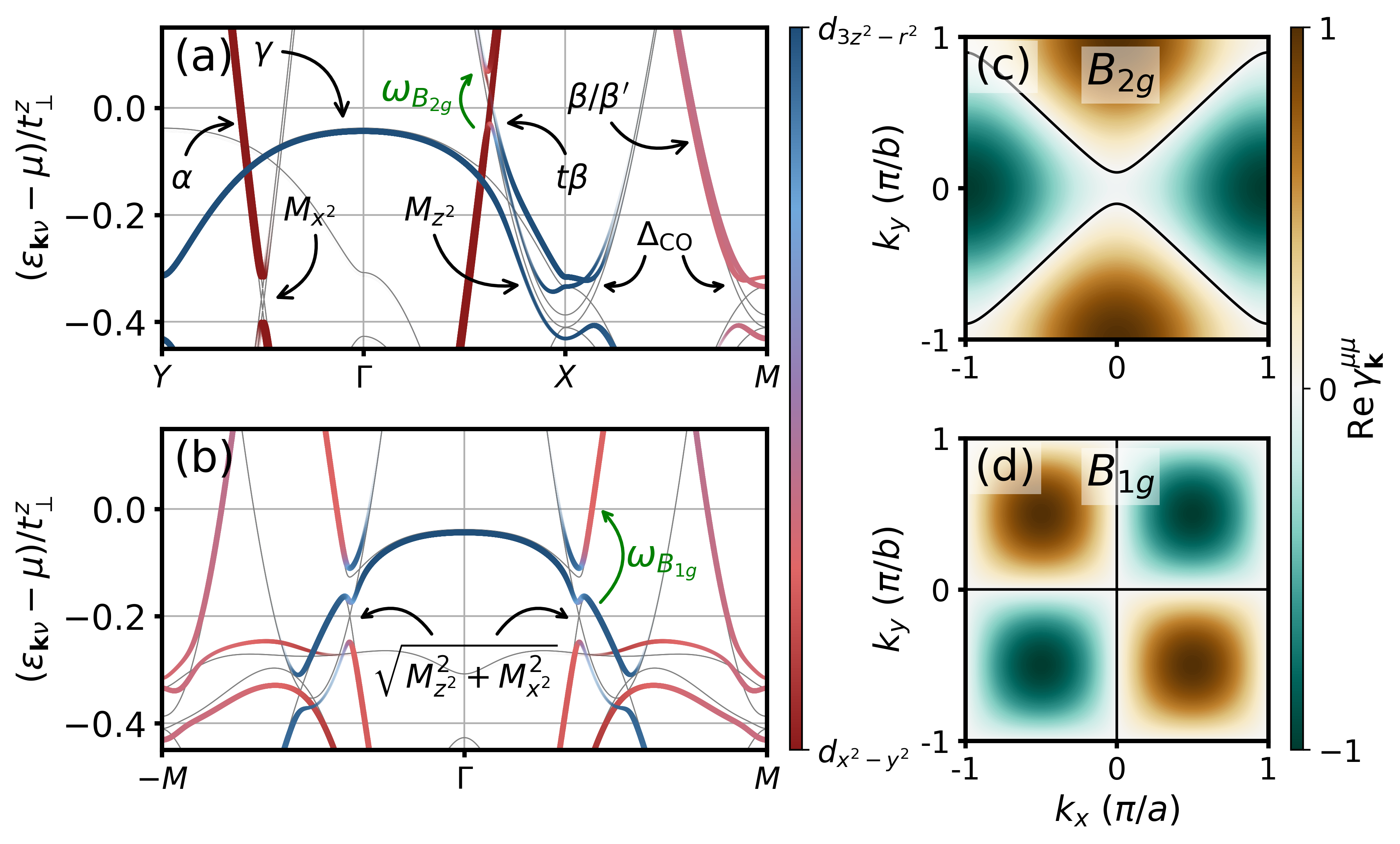}
\caption{ Multiorbital electronic structure and symmetry-resolved Raman vertices in the spin-density-wave (SDW) state of bilayer nickelates with pre-existing $Am2m$ charge order. Due to orthorhombicity the vertices are not $C_4$ symmetric.  (a,b) Calculated SDW-reconstructed electronic band structure along the high-symmetry paths $\boldsymbol{Y}-\boldsymbol{\Gamma}-\boldsymbol{X}-\boldsymbol{M}$ and $-\boldsymbol{M}-\boldsymbol{\Gamma}-\boldsymbol{M}$, respectively. The color scale indicates the relative orbital contributions of Ni $d_{3z^2-r^2}$ (blue) and $d_{x^2-y^2}$ (red) orbitals, while thin gray lines show the corresponding bands in the absence of SDW order, retaining the underlying charge disproportionation. Arrows identify the $\alpha$, $\gamma$, $\beta$, $\beta'$, and $t\beta$ bands, as well as the characteristic energy scales associated with the orbital-dependent SDW gaps, $M_{z^2}$ and $M_{x^2}$, the dominant $B_{2g}$ ($\omega_{B_{2g}}$) and $B_{1g}$ ($\omega_{B_{1g}}$) peak energies, the charge-order-induced splitting $\Delta_{\text{CO}}$, and the multiorbital SDW gap $\sqrt{M_{z^2}^2+M_{x^2}^2}$. 
Results are shown for $U/t_\perp^z=1.98$, $J_H=U/4$, and $T/T_{\text{SDW}}\simeq 0.76$. 
(c,d) {Calculated momentum dependence of the real part of the orbital-resolved Raman vertex $\gamma_{\boldsymbol{k}}^{\mu\mu}$ for $\mu=d_{3z^2-r^2}$ in the $B_{2g}$ and $B_{1g}$ symmetry channels, respectively. Black contours indicate zeros of the Raman vertex, separating regions of opposite sign. 
The distinct momentum dependence of the two vertices illustrates how the Raman polarization geometries probe different regions of the reconstructed electronic structure.} 
The green arrows indicate the characteristic energy scattering processes.
}
\label{fig:model}
\end{center}
\end{figure}

The onset of SDW order reconstructs the electronic band structure through folding by the magnetic ordering vector $\boldsymbol{Q}_Y=(0,\pi)$ and the opening of gaps at crossings between the original and folded bands. Figures~\ref{fig:model}(a) and (b) show the resulting band structure in the coexistence phase of the pre-existing $Am2m$ charge order and spin-large spin-low stripe SDW order. The colors indicate the relative contributions of the Ni $d_{x^2-y^2}$ (red) and $d_{3z^2-r^2}$ (blue) orbitals, while the thin gray lines represent the corresponding (folded) bands in the absence of magnetic order. Although the overall band dispersions remain largely unchanged, SDW order produces several characteristic gap openings whose magnitudes depend on the orbital composition of the participating states. 

In particular, along the $\boldsymbol{Y}-\boldsymbol{\Gamma}-\boldsymbol{X}-\boldsymbol{M}$ path [Fig.~\ref{fig:model}(a)], the SDW-induced gaps can be classified according to the orbital character of the intersecting bands. First, the predominantly $d_{x^2-y^2}$-derived electron-like $\alpha$ band is folded from the vicinity of $\boldsymbol{\Gamma}$ towards $\boldsymbol{Y}$, where it intersects its original counterpart. The resulting intraband $\alpha$--$\alpha$ hybridization opens a gap characterized by the $d_{x^2-y^2}$ orbital-dependent magnetic order parameter, $|M_{x^2}|$. A second class of gaps involves the $\beta$ and $\beta'$ bands, which contain contributions from both Ni $e_g$ orbitals. The additional $\beta'$ band arises from the enlarged four-Ni unit cell associated with the orthorhombic structure. Upon folding by $\boldsymbol{Q}_Y$, these bands are mapped from the vicinity of $\boldsymbol{M}$ towards $\boldsymbol{X}$ and intersect the $\alpha$- and $\gamma$-derived states. The $\beta/\beta'$--$\alpha$ crossings generate SDW gaps near the Fermi level, predominantly involving $d_{x^2-y^2}$ orbital character, whereas the $\beta/\beta'$--$\gamma$ crossings produce additional gaps within the occupied electronic states \cite{Braz2026DensityWaves}. In the latter case, the predominantly $d_{3z^2-r^2}$ character of the $\gamma$ band selects the corresponding orbital component of the folded $t\beta$ band, giving rise to a characteristic gap scale associated with $|M_{z^2}|$.

The pre-existing charge disproportionation further modifies this reconstruction. In particular, the charge-order potential $\Delta_{\text{CO}}$ lifts the near-degeneracy of the $\beta$ and $\beta'$ bands and modifies their intersections with the SDW-folded states, thereby enhancing the visibility of the $t\beta$ band near the Fermi level. As discussed in Appendix~\ref{appendix:co}, the resulting electronic reconstruction differs from that obtained in the absence of the high-temperature charge order. The predominantly $d_{x^2-y^2}$ character of the low-energy Fermi surface, together with the appearance of the folded $t\beta$ feature, is consistent with available ARPES observations \cite{AuYeung2025UniversalNickelates,Braz2026DensityWaves}.

An important feature of the reconstruction along $\boldsymbol{Y}-\boldsymbol{\Gamma}-\boldsymbol{X}-\boldsymbol{M}$ is that the relevant SDW-induced gaps are associated with a single orbital component, despite the multiorbital character of the underlying electronic structure. This contrasts with the situation along the $\boldsymbol{\Gamma}-\boldsymbol{M}$ direction [Fig.~\ref{fig:model}(b)], where the intersecting bands exhibit substantial admixture of both Ni $e_g$ orbitals. The SDW coupling between these strongly hybridized states produces a characteristic gap scale of approximately $\sqrt{M_{z^2}^2+M_{x^2}^2}$, reflecting the combined contributions of the two orbital-dependent magnetic order parameters.
The coexistence of predominantly intraorbital and strongly hybridized SDW gaps is particularly relevant for the electronic Raman response. {Because the $B_{1g}$ and $B_{2g}$ Raman vertices have different momentum and orbital dependences [Figs.~\ref{fig:model}(c) and (d)], they weight electronic transitions involving these distinct gap scales differently. As we show below, this provides a microscopic explanation for the different characteristic energies observed in the two Raman symmetry channels. Observe also that the Raman response probes transitions between occupied and unoccupied quasiparticle states rather than the positions of individual band gaps. The SDW-induced reconstruction generates such transitions near avoided band crossings, as indicated by the green arrows in Figs. ~\ref{fig:model}(a) and (b). Their characteristic energies and spectral weights depend on the reconstructed dispersions, orbital composition, and Raman matrix elements.}

\begin{figure*}[t]
\begin{center}
\includegraphics[width=1.0\textwidth]{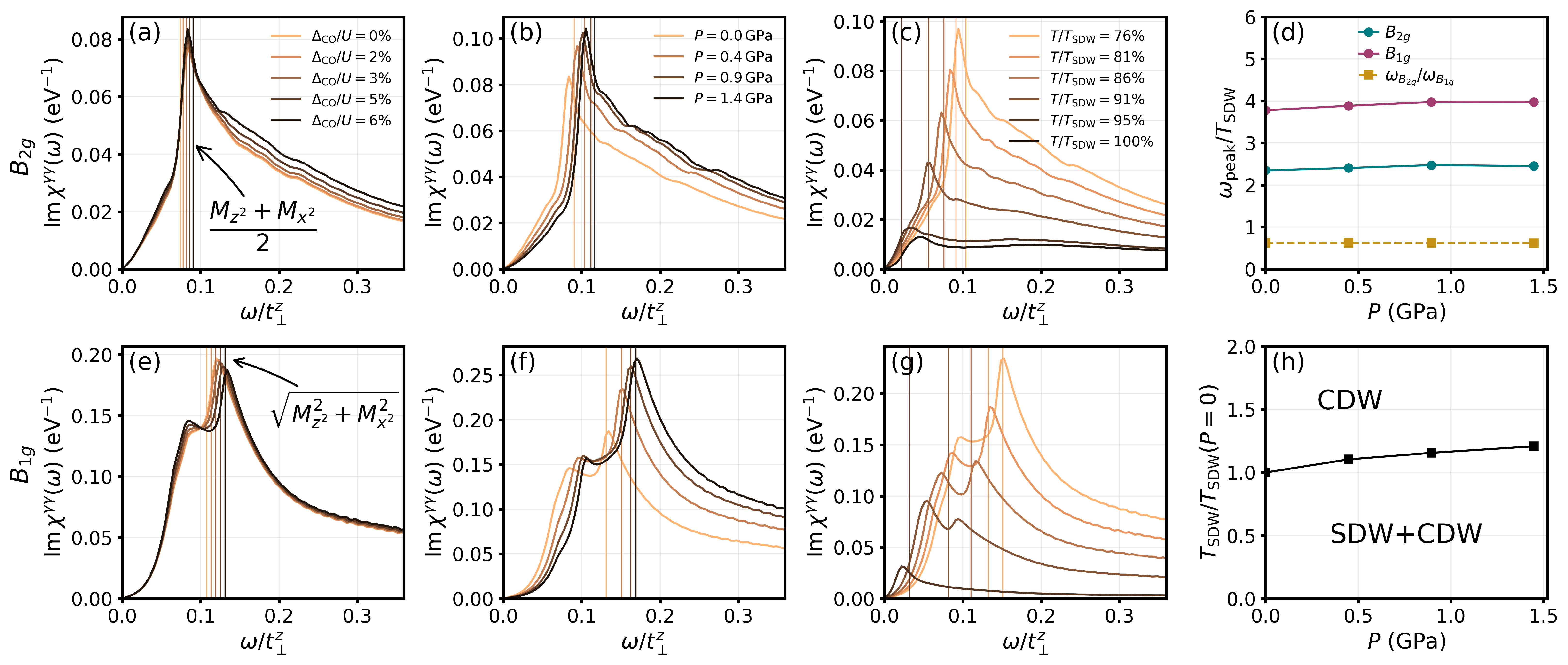}
\caption{ Calculated symmetry-resolved electronic Raman response in the spin-density-wave (SDW) state of bilayer nickelates with pre-existing charge order in the $Am2m$-crystallographic phase. The imaginary part of the Raman susceptibility, $\text{Im},\chi^{\gamma\gamma}(\omega)$, is shown for the $B_{2g}$ (upper row) and $B_{1g}$ (lower row) symmetry channels. (a,e) Evolution of the Raman spectra with increasing charge-order potential, $\Delta_{\text{CO}}/U$, at fixed reduced temperature $T/T_{\text{SDW}}=0.76$. (b,f) Pressure dependence of the Raman response for $\Delta_{\text{CO}}/U=0.06$ and $T/T_{\text{SDW}}=0.76$. (c,g) Temperature dependence at fixed pressure $P=0.4$ GPa and $\Delta_{\text{CO}}/U=0.04$, illustrating the progressive suppression of the SDW-related spectral features upon approaching and crossing the magnetic transition. Frequencies are expressed in units of the interlayer hopping amplitude $t_\perp^z$. Thin vertical lines indicate the characteristic SDW energy scales associated with the corresponding model parameters. (d) Pressure evolution of the dominant Raman peak positions, $\omega_{B_{2g}}$ and $\omega_{B_{1g}}$, normalized by the pressure-dependent SDW transition temperature, together with their ratio $\omega_{B_{2g}}/\omega_{B_{1g}}$. (h) Pressure dependence of the SDW transition temperature, normalized by its ambient-pressure value, showing the coexistence of SDW and charge order below $T_{\text{SDW}}$ and the persistence of the charge-disproportionated phase above the magnetic transition. All calculations reported use $U/t_\perp^z=1.98$ and $J_H=U/4$.
}
\label{fig:raman}
\end{center}
\end{figure*}

\section{Raman signatures of a multiorbital SDW} 
Motivated by the agreement of the calculated electronic structure and magnetic ordering pattern with available ARPES and local magnetic probe measurements, we now turn to the electronic Raman response. Our aim is to establish how the multiorbital reconstruction induced by SDW order manifests itself in the symmetry-resolved Raman spectra and, in particular, whether it can account for the distinct spectral features observed experimentally in the $B_{1g}$ and $B_{2g}$ channels. A detailed derivation of the coupling between the incoming light beam and the self-consistent mean-field Hamiltonian is presented in Appendix \ref{appendix:e-l_coupling}.
In particular, we consider the experiment where an incoming beam with magnetic vector potential $\bbs A$ and momentum $-e\bbs A/c$ scatters with a mean-field quasiparticle with momentum $\bbs k$. This scattering event shifts the quasiparticle momentum to $\bbs k'=\bbs k-e\bbs A/c$.
Expanding the effective electron-light Hamiltonian into powers of the vector potential retakes the original dependence of the expansion on the quasiparticle momentum $\bbs k$.
At lowest order in $\bbs A$, one obtains the so-called multiorbital Raman ``vertex'', which resembles an ``effective-mass'' term
\begin{equation}
\gamma^{\mu\eta}_{\bbs k}
=
\frac{e^2}{2c^2}\sum_{ij}A_i \frac{\partial^2 h_{\mu\eta}(\bbs k)}{\partial k_i\partial k_j}A_j,
\label{eq:raman_vertex_main}
\end{equation}
where the spin channel is unimportant, because it only contains momentum-independent contributions in orbital space. The Raman response function comes from treating the electron-light coupling as a quadratic source field.
A diagrammatic expansion from the path-integral formalism is fully derived in Appendix~\ref{appendix:derivation} and results in a bubble diagram dressed with the self-consistently determined mean-field Green's functions. 
{The Raman vertices are evaluated directly from the momentum derivatives of the full multiorbital tight-binding Hamiltonian in the orbital basis, rather than approximated by single-band trigonometric form factors. Consequently, both their orbital and momentum dependence, including the effects of orthorhombicity, are retained. 
The transformation to the quasiparticle basis is then accounted for through the eigenvectors entering the Raman susceptibility.}

{Experiments can tune $\bbs A$ by polarizing the incoming beam to desired spatial directions.
It is convenient to decompose the different possibilities of $\bbs A$ onto symmetry channels of the underlying crystal structure.
However, both $Amam$ and $Am2m$ space groups are non-symmorphic and its irreducible representations are quite complicated.
A more natural choice is the in-plane tetragonal-structure $D_{4h}$ point-group, with irreducible representations $A_{1g}$, $B_{1g}$, and $B_{2g}$.
In Appendix \ref{appendix:e-l_coupling}, we detail the decomposition of symmetry channels in terms of the vector potential $\boldsymbol{A}$.}
Indeed, this is a reasonable choice of decomposition, because the in-plane orthorhombic lattice parameters in low-pressure bilayer nickelates differ by only $2\%$ \cite{Xu2026,Plokhikh2025SpinDensityNickelate}.
As shown in Fig.~\ref{fig:model}(c) and (d), the calculation of Eq.~\eqref{eq:raman_vertex_main} results in a Raman vertex that is tetragonal in the $B_{1g}$ channel and slightly orthorhombic in the $B_{2g}$ channel. 
{Also, the $B_{1g}$ and $B_{2g}$ labels used here follow the polarization conventions of the experimental crystallographic coordinate system. 
They should not be directly identified with the corresponding form factors in an unfolded one-Ni Brillouin zone, whose in-plane axes are rotated relative to the crystallographic description by 45$^\circ$, as illustrated in Fig.~\ref{fig:bz_conventions} of Appendix \ref{appendix:e-l_coupling}. 
This distinction is important when comparing our Raman vertices with those obtained from simplified one-Ni models.}

Figure~\ref{fig:raman} presents the calculated electronic Raman response of La$_3$Ni$_2$O$_7$ in the $B_{2g}$ [Figs.~\ref{fig:raman}(a)--(c)] and $B_{1g}$ [Figs.~\ref{fig:raman}(e)--(g)] symmetry channels, with frequencies expressed in units of the interlayer hopping amplitude $t_\perp^z$. We focus on these two channels because the fully symmetric $A_{1g}$ response is strongly affected by long-range Coulomb screening in metallic systems \cite{Dierker1983ElectronicRaman,Devereaux2007InelasticLightScattering}, which requires a separate treatment. For completeness, we present the unscreened $A_{1g}$ response in Appendix~\ref{appendix:A1g}, where its main spectral features are found to resemble those of the $B_{2g}$ channel.

{We first examine the influence of the pre-existing charge disproportionation on the Raman spectra. As shown in Figs.~\ref{fig:raman}(a) and (e), increasing the charge-order potential $\Delta_{\text{CO}}/U$ enhances the Raman intensity in both symmetry channels and produces a redistribution of spectral weight towards higher energies. This evolution reflects the influence of charge disproportionation on the self-consistent magnetic order parameters and the resulting reconstruction of the electronic bands, as discussed in Appendix~\ref{appendix:co}. The enhancement is particularly pronounced in the $B_{1g}$ channel, where a broad high-energy feature develops and eventually dominates the response. These results demonstrate that the pre-existing charge order plays an important role in shaping the electronic Raman spectra of the SDW state.}

The pressure dependence, shown in Figs.~\ref{fig:raman}(b) and (f), reveals a distinct evolution of the two polarization channels. In the $B_{2g}$ response, increasing pressure suppresses the overall intensity and shifts the dominant peak towards lower energies. This behavior correlates with the reduction of the spin-density modulation and charge disproportionation obtained in the self-consistent calculations [Figs.~\ref{fig:op}(a) and (b)]. A similar tendency is observed for the lower-energy shoulder of the $B_{1g}$ response. In contrast, the dominant higher-energy $B_{1g}$ feature remains comparatively stable in frequency, although its intensity decreases and its spectral profile becomes broader with increasing pressure. This difference suggests that the two Raman channels probe distinct aspects of the SDW-induced electronic reconstruction.
The temperature evolution provides further insight into the origin of these features. As shown in Figs.~\ref{fig:raman}(c) and (g), the characteristic SDW-related peaks progressively weaken upon approaching $T_{\text{SDW}}$, reflecting the suppression of the magnetic order parameters. Above the transition, the pronounced SDW-related structures disappear, while a residual electronic continuum remains in the $B_{2g}$ channel. Within the present calculation, the latter originates from interband transitions already allowed in the charge-disproportionated normal state. The much weaker normal-state $B_{1g}$ response reflects the different momentum and orbital structure of its Raman vertex. As discussed in Appendix~\ref{appendix:co}, the orbital dependence of the $B_{2g}$ vertex favors particular interorbital transitions, providing a natural explanation for the different normal-state responses of the two channels. We note, however, that the present mean-field treatment does not account for short-range magnetic correlations or fluctuations above $T_{\text{SDW}}$. While the charge-disproportionated normal state supports a residual electronic continuum, particularly in the $B_{2g}$ channel, the persistence of experimentally observed SDW-like spectral features above the transition may require additional fluctuation effects beyond our approach.

A central result of our calculations is the emergence of two distinct characteristic Raman energy scales. Throughout the parameter range investigated, the dominant $B_{2g}$ peak occurs at a lower frequency than the principal $B_{1g}$ feature. The vertical lines in Figs.~\ref{fig:raman}(a)--(c) indicate the effective single-orbital scale $(M_{z^2}+M_{x^2})/2$, while those in Figs.~\ref{fig:raman}(e)--(g) indicate the multiorbital scale $\sqrt{M_{z^2}^2+M_{x^2}^2}$. Remarkably, these independently determined combinations of the orbital-dependent SDW order parameters track the characteristic spectral features in the corresponding symmetry channels. In particular, the lower-energy $B_{2g}$ peak follows the average single-orbital scale, whereas the higher-energy $B_{1g}$ feature is associated with the combined multiorbital scale. The additional lower-energy shoulder in the $B_{1g}$ response is likewise connected to transitions involving the predominantly intraorbital SDW gaps.

{The microscopic origin of these two energy scales can be understood from the reconstructed electronic structure in Fig.~\ref{fig:model}. Along the $\boldsymbol{Y}-\boldsymbol{\Gamma}-\boldsymbol{X}-\boldsymbol{M}$ path [Fig.~\ref{fig:model}(a)], the SDW opens gaps involving bands with relatively weak interorbital hybridization, characterized predominantly by the individual order parameters $M_{z^2}$ and $M_{x^2}$. These regions are strongly weighted by the $B_{2g}$ Raman vertex [Fig.~\ref{fig:model}(c)]. By contrast, along the $\boldsymbol{\Gamma}-\boldsymbol{M}$ direction [Fig.~\ref{fig:model}(b)], the SDW couples states with substantial contributions from both Ni $e_g$ orbitals, producing a characteristic energy scale associated with $\sqrt{M_{z^2}^2+M_{x^2}^2}$. The $B_{1g}$ vertex [Fig.~\ref{fig:model}(d)] preferentially weights these regions of the Brillouin zone. Thus, the separation between the dominant Raman peaks originates from the interplay of orbital-dependent SDW gaps, interorbital hybridization, and the momentum selectivity of the Raman vertices.
Another prominent feature is the substantially stronger Raman response in the $B_{1g}$ channel. Within the present calculation, its enhanced intensity is predominantly associated with the broad higher-energy feature that develops below $T_{\text{SDW}}$. The lower-energy shoulder, by comparison, has an intensity comparable to the corresponding $B_{2g}$ response. This indicates that the difference between the two channels is not simply an overall enhancement of the Raman coupling, but is closely connected to the additional electronic transitions involving strongly hybridized multiorbital states. The reconstructed band structure provides these transitions with substantial Raman matrix elements and available spectral weight, resulting in a more pronounced $B_{1g}$ response.}

{The different electronic dispersions contributing to the two symmetry channels also provide an interpretation of their distinct spectral line shapes. The dominant $B_{2g}$ response is associated with transitions between relatively dispersive bands separated by predominantly intraorbital SDW gaps, giving rise to a comparatively sharp spectral feature. In contrast, the higher-energy $B_{1g}$ response receives substantial contributions from transitions involving a flatter band and a more dispersive strongly hybridized band near the Fermi level. 
These transitions extend over a broader range of excitation energies, producing a less sharply defined spectral maximum. The momentum dependence of the Raman vertices further enhances this distinction by weighting the corresponding transitions differently. Thus, the characteristic line shapes reflect not only the SDW gap magnitudes but also the dispersions and orbital composition of the reconstructed bands.}

To quantify the pressure evolution, Fig.~\ref{fig:raman}(d) compares the dominant Raman peak frequencies normalized by the pressure-dependent SDW transition temperature, $\omega_{\text{peak}}/T_{\text{SDW}}$. 
The $B_{1g}$ ratio approaches a plateau near $\omega_{\text{peak}}^{B_{1g}}/T_{\text{SDW}}\simeq4$, reflecting the comparatively weak pressure dependence of the higher-energy peak and the saturation of $T_{\text{SDW}}$ [Fig.~\ref{fig:op}(h)]. 
Similarly, the normalized $B_{2g}$ peak frequency approaches $\omega_{\text{peak}}^{B_{2g}}/T_{\text{SDW}}\simeq2.5$ at higher pressures. 
The resulting reduction of $\omega_{\text{peak}}^{B_{2g}}/\omega_{\text{peak}}^{B_{1g}}$ highlights the increasing separation between the two characteristic Raman energy scales. 
The approximate magnitude of the $B_{1g}$ ratio is consistent with recent experimental observations \cite{Zhang2025IdentifyingStructureLa3Ni2O7}. 
We emphasize, however, that our phenomenological description of pressure through the crystal-field splitting is expected to be most reliable below approximately 3 GPa. 
At higher pressures, the saturation of the crystal-field splitting makes pressure-induced modifications of the hopping amplitudes increasingly relevant \cite{Braz2026DensityWaves}.

Our results provide a unified electronic interpretation of several characteristic features reported in Raman experiments on bilayer nickelates \cite{He2026AnisotropicNickelate,Gim2026SpectroscopicNickelate,Zhou2026UnconventionalPressureEvolution}. 
First, the larger $B_{1g}$ intensity arises predominantly from transitions involving strongly hybridized multiorbital states. 
Second, the experimentally observed hierarchy of peak energies, $\omega_{\text{peak}}^{B_{1g}}>\omega_{\text{peak}}^{B_{2g}}$, follows naturally from the distinct SDW-related energy scales selected by the two Raman vertices. 
Third, the coexistence of a relatively sharp lower-energy feature and a broader higher-energy response provides a microscopic electronic mechanism for the different spectral profiles observed in the two symmetry channels. 
Finally, the similar pressure evolution of the peaks reflects the sensitivity of predominantly intraorbital and strongly hybridized electronic transitions to changes in the underlying band structure.
The separation between the two Raman energy scales also offers a possible means of constraining the relative magnitudes of the orbital-dependent magnetic order parameters. 
Assuming that the experimentally observed peak energies can be identified with the corresponding characteristic scales, and using the hierarchy $M_{z^2}>M_{x^2}$ obtained in our calculations, the experimental data of Ref.~\cite{He2026AnisotropicNickelate} suggest a substantially larger $d_{3z^2-r^2}$ magnetic contribution, with an estimated ratio $M_{x^2}/M_{z^2}\sim0.15$. 
This estimate should be regarded as model-dependent, since the precise Raman peak positions also depend on the electronic dispersions, orbital hybridization, and Raman matrix elements. 
Nevertheless, the inferred predominance of the $d_{3z^2-r^2}$ orbital is compatible with the importance of magnetic exchange processes emphasized in local-moment descriptions of neutron scattering \cite{Xie_StrongInterlayerMagneticExchangeCoupling_2024,Zhou2026SpinFluctuationsNickelates,Chen2026NatureMagnetismNickelate}, RIXS \cite{Gupta2025SpinStripeNickelate}, and Raman spectroscopy \cite{Gim2026SpectroscopicNickelate}.

Our results demonstrate that the characteristic symmetry dependence of the Raman spectra can arise from electronic particle--hole excitations in a multiorbital SDW state. The pre-existing $Am2m$ charge disproportionation, the subsequent development of spin-modulated stripe order, and the orbital selectivity of the Raman vertices provide a consistent microscopic framework for understanding the electronic reconstruction observed by ARPES and the magnetic inhomogeneity indicated by local probes. Importantly, although the present calculation does not explicitly include two-magnon Raman processes and therefore cannot exclude their contribution to the experimental spectra, it establishes that the principal separation of Raman energy scales and their contrasting pressure dependence do not require distinct magnetic collective excitations. More generally, our findings demonstrate the sensitivity of symmetry-resolved electronic Raman spectroscopy to orbital-dependent density-wave reconstruction in multiorbital correlated materials.

\section{Conclusions}
In summary, we have investigated the symmetry-resolved electronic Raman response of bilayer nickelates in the spin-density-wave (SDW) state using a multiorbital itinerant model that incorporates the pre-existing charge disproportionation associated with the low-pressure $Am2m$ crystal structure. Our self-consistent calculations demonstrate that the onset of stripe magnetic order on top of the charge-disproportionated electronic state generates a modulation of the local magnetic moments and reconstructs the electronic bands through orbital-dependent SDW gap openings. The interplay between charge disproportionation, orbital hybridization, and magnetic order provides a microscopic framework for understanding the electronic and magnetic properties of low-pressure La$_3$Ni$_2$O$_7$.

Our central finding is that the reconstructed multiorbital electronic structure naturally produces distinct characteristic energy scales in the $B_{1g}$ and $B_{2g}$ Raman channels. The lower-energy $B_{2g}$ response is predominantly associated with electronic transitions across SDW gaps involving weakly hybridized orbital states, characterized approximately by the energy scale $(M_{z^2}+M_{x^2})/2$. In contrast, the dominant higher-energy $B_{1g}$ feature originates from transitions involving strongly hybridized multiorbital bands and follows a characteristic scale of approximately $\sqrt{M_{z^2}^2+M_{x^2}^2}$. The different momentum and orbital selectivity of the Raman vertices thus provides a natural explanation for the separation of the two spectral features and their distinct line shapes.

We further demonstrate that the pre-existing charge disproportionation substantially enhances the Raman response and influences the redistribution of spectral weight in the SDW state. The calculated pressure and temperature evolution reproduces several key experimental trends, including the stronger $B_{1g}$ response, the higher energy of its dominant peak, and the contrasting pressure dependence of the two symmetry channels. In particular, the $B_{2g}$ peak softens relative to the SDW transition temperature under pressure, whereas the dominant $B_{1g}$ feature remains comparatively robust. These results highlight that the characteristic electronic excitation energies and the magnetic ordering temperature need not evolve in parallel, reflecting the multiorbital nature of the underlying SDW state.

Our findings also provide a perspective on the ongoing discussion concerning the microscopic origin of the Raman features observed in bilayer nickelates. While collective magnetic excitations, including two-magnon processes, may contribute to the experimental response, our calculations establish that the principal symmetry-dependent spectral features can arise naturally from electronic particle--hole excitations in the reconstructed SDW state. In addition, we find that polarization-resolved Raman spectroscopy provides a sensitive probe of orbital-dependent magnetic order and multiorbital electronic reconstruction, offering a route to disentangling the intertwined charge, spin, and orbital degrees of freedom in unconventional superconducting materials.

\begin{acknowledgments}
We thank Meng Wang, Xianxin Wu, Marius Scholten, Steffen B\"otzel and Frank Lechermann for fruitful discussions. L.B.B. acknowledges financial support from the grants 2023/14902-8 and 2025/17852-7, São Paulo Research Foundation (FAPESP).
L.G.G.V.D.S. acknowledges financial support by Brazilian agencies CNPq (Grant. No. 312622/2023-6), and FAPESP (Grant no. 2022/15453-0). 
G.H. acknowledges support from the National Natural Science Foundation of China (Grant No. 12474473).
\end{acknowledgments}

\appendix
\section{Path-integral formulation of multiorbital density waves}
\label{appendix:theory}
Here, based on Ref. \cite{Hoyer2018RoleFluctuations}, we detail the theory for spin and charge order parameters as probed for in our Hartree-Fock calculations.
We consider the multiorbital Hamiltonian
\begin{align}
H = H_0 + H_{\text{int}},
\end{align}
with the quadratic part $H_0$ given by Eqs.~\eqref{eq:H0_sublattice} and the and quartic part $H_{\text{int}}$ by the Hubbard-Kanamori Hamiltonian
\begin{equation}
\begin{split}
H_{\mathrm{int}}
=
\frac{1}{4}
\sum_{\boldsymbol{q}}
&\sum_{\mu\eta\alpha\beta}
\Big[
(U_c)^{\mu\eta}_{\alpha\beta}
n_{-\boldsymbol{q}\mu\eta}
n_{\boldsymbol{q}\alpha\beta} \\
&\qquad-
(U_s)^{\mu\eta}_{\alpha\beta}
\boldsymbol{S}_{-\boldsymbol{q}\mu\eta}
\cdot
\boldsymbol{S}_{\boldsymbol{q}\alpha\beta}
\Big],
\label{eq:Kanamori_nc_form}
\end{split}
\end{equation}
where the electron density $n_{\boldsymbol{q}\mu\eta}
=
\sum_{\boldsymbol{k}\sigma}
c^{\dagger}_{(\boldsymbol{k}+\boldsymbol{q})\mu\sigma}
c_{\boldsymbol{k}\eta\sigma}$
and spin density
$\boldsymbol{S}_{\boldsymbol{q}\mu\eta}
=
\sum_{\boldsymbol{k}\sigma\sigma'}
c^{\dagger}_{(\boldsymbol{k}+\boldsymbol{q})\mu\sigma}
\boldsymbol{\sigma}_{\sigma\sigma'}
c_{\boldsymbol{k}\eta\sigma'}$
are defined, with
$\boldsymbol{\sigma}=(\sigma^x,\sigma^y,\sigma^z)$ denoting the Pauli vector.
Interaction tensors given by \cite{kemperSensitivitySuperconductingState2010}
\begin{align}
(U_s)^{\mu\eta}_{\alpha\beta}
=
\begin{cases}
U, & \mu=\eta=\alpha=\beta, \\
J_H, & \mu=\eta\neq\alpha=\beta, \\
J_H', & \mu=\alpha\neq\eta=\beta, \\
U', & \mu=\beta\neq\eta=\alpha, \\
0, & \text{otherwise}.
\end{cases}\label{eq:Us}
\end{align}
\begin{align}
(U_c)^{\mu\eta}_{\alpha\beta}
=
\begin{cases}
U, & \mu=\eta=\alpha=\beta, \\
- J_H + 2U', & \mu=\eta\neq\alpha=\beta, \\
J_H', & \mu=\alpha\neq\eta=\beta, \\
- U' + 2J_H, & \mu=\beta\neq\eta=\alpha, \\
0, & \text{otherwise},\label{eq:Uc}
\end{cases}
\end{align}
where it is implicit that the index combination must be constrained to be onsite.

Introducing Grassmann fields
$c_{\boldsymbol{k}\mu\sigma}(\tau)$ and
$c^{\dagger}_{\boldsymbol{k}\mu\sigma}(\tau)$,
the action for the Hamiltonian $H$ is
\begin{align}
S
=
\int_0^\beta d\tau
\Bigg[
\sum_{\boldsymbol{k}\mu\eta\sigma}
c^{\dagger}_{\boldsymbol{k}\mu\sigma}
(\partial_\tau \delta_{\mu\eta} + \xi^{\mu\eta}_{\boldsymbol{k}})
c_{\boldsymbol{k}\eta\sigma}
+
H_{\mathrm{int}}(\tau)
\Bigg].
\end{align}
Explicitly,
\begin{align}
S
=
\int_0^\beta d\tau
\sum_{\boldsymbol{k}\mu\eta\sigma}
c^{\dagger}_{\boldsymbol{k}\mu\sigma}
(\partial_\tau \delta_{\mu\eta} + \xi^{\mu\eta}_{\boldsymbol{k}})
c_{\boldsymbol{k}\eta\sigma}
+
S_{\mathrm{int}},
\end{align}
with
\begin{equation}
\begin{split}
    S_{\mathrm{int}}
=
\frac{1}{4N}
\int_0^\beta d\tau
\sum_{\boldsymbol{q}}
&\sum_{\mu\eta\alpha\beta}
\Big[
(U_c)^{\mu\eta}_{\alpha\beta}
\, n_{-\boldsymbol{q}\mu\eta}
\, n_{\boldsymbol{q}\alpha\beta}\\
&-
(U_s)^{\mu\eta}_{\alpha\beta}
\, \boldsymbol{S}_{-\boldsymbol{q}\mu\eta}
\cdot
\boldsymbol{S}_{\boldsymbol{q}\alpha\beta}
\Big].
\end{split}
\end{equation}
$\boldsymbol{\sigma}$ denotes the Pauli vector.
This action is quartic in fermionic fields. 
We then introduce charge and spin auxiliary fields
$\phi_{\boldsymbol{q}\mu\eta}(\tau)$ and
$\boldsymbol{M}_{\boldsymbol{q}\mu\eta}(\tau)$.
The Hubbard-Stratonovich transformation can be schematically represented as \cite{Hoyer2018RoleFluctuations}
\begin{align}
e^{\frac{1}{4}\sum_q\hat{\bbs b}_{-q} \hat{V}\hat{\bbs b}_q} = \int \mathcal{D}\hat{\bbs x} e^{ \sum_q ( -\hat{\bbs x}_{-q} \cdot \hat{V}^{-1} \hat{\bbs x}_{q} + \frac{1}{2}\hat{\bbs x}_{-q}\cdot\hat{\bbs b}_{q} + \frac{1}{2}\hat{\bbs b}_{-q}\cdot\hat{\bbs x}_{q} )}.
\label{eq:HS_int}
\end{align}
We define, for each charge and spin Hubbard-Stratonovich field,
\begin{align}
&\hat{\bbs x}_q \rightarrow \hat{\phi}_q,\; \bbs b_q \rightarrow n_{q\mu\eta},\; V \rightarrow \hat{U}_c \\
&\hat{\bbs x}_q \rightarrow -\hat{\bbs M}_q,\; \bbs b_q \rightarrow \bbs{S}_{\bbs{q}\mu\eta},\; V \rightarrow -\hat U_s,
\label{eq:HS_transform}
\end{align}
resulting in a final action
\begin{equation}
    S=S_{\text{ferm}}+S_{\text{bos}}
\label{eq:action}
\end{equation}
where the fermionic part can be rewritten by conveniently relabeling the momenta $\bbs q$ as
\begin{equation}
\begin{split}
S_{\text{ferm}}
&=
\int_0^\beta d\tau
\sum_{\boldsymbol{k} \boldsymbol{k}'\mu\eta\sigma\sigma'}
c^{\dagger}_{\boldsymbol{k}\mu\sigma}
\Big[
(\partial_\tau \delta_{\boldsymbol{k}\boldsymbol{k}'}\delta_{\mu\eta}+
\xi^{\mu\eta}_{\boldsymbol{k}} \delta_{\boldsymbol{k}\boldsymbol{k}'}\\
&+
\phi_{(\boldsymbol{k}-\boldsymbol{k}')\mu\eta})\delta_{\sigma\sigma'}
-
\boldsymbol{M}_{(\boldsymbol{k}-\boldsymbol{k}')\mu\eta}
\cdot
\boldsymbol{\sigma}_{\sigma\sigma'}
\Big]
c_{\boldsymbol{k}'\eta\sigma'}.
\end{split}
\label{eq:S_ferm}
\end{equation}
with a bosonic contribution
\begin{equation}
\begin{split}
S_{\text{bos}}=
\frac{1}{N}
\int_0^\beta d\tau
\sum_{\boldsymbol{q}\mu\eta\alpha\beta}
&\Big[
-\phi_{-\boldsymbol{q}\mu\eta}
(U_c^{-1})^{\mu\eta}_{\alpha\beta}
\phi_{\boldsymbol{q}\alpha\beta}\\
&+
\boldsymbol{M}_{-\boldsymbol{q}\mu\eta}
\cdot
(U_s^{-1})^{\mu\eta}_{\alpha\beta}
\boldsymbol{M}_{\boldsymbol{q}\alpha\beta}
\Big].
\end{split}
\label{eq:S_bos}
\end{equation}

Since the fermionic action is quadratic, we integrate out the fermions exactly performing a Gaussian Berezin integration such that the partition function becomes
\begin{align}
Z
=
\int
\mathcal{D}[\hat{\phi},\hat{\boldsymbol{M}}]
\, e^{-S_{\mathrm{eff}}[\hat{\phi},\hat{\boldsymbol{M}}]},\label{eq:Z}
\end{align}
with
\begin{equation}
\begin{split}
S_{\mathrm{eff}}
&=
\frac{1}{N}
\int_0^\beta d\tau
\sum_q
\Big[
-\hat{\phi}_{-\boldsymbol{q}} \hat{U}_c^{-1} \hat{\phi}_{\boldsymbol{q}}
+
\hat{\boldsymbol{M}}_{-\boldsymbol{q}}\cdot \hat{U}_s^{-1} \hat{\boldsymbol{M}}_{\boldsymbol{q}}
\Big] \\
&-
\mathrm{Tr}\ln
\Big[
\hat{G}^{-1}(\phi,\boldsymbol{M})
\Big],
\label{eq:Seff_Kanamori}
\end{split}
\end{equation}
where the orbital indices have been dropped to give room for a matrix notation and the inverse Green's function reads
\begin{align}
\hat{G}^{-1}_{\boldsymbol{k} \boldsymbol{k}'}
=
\hat{\partial}_\tau \delta_{\boldsymbol{k}\boldsymbol{k}'}
+
\hat{h}_{\boldsymbol{k}\boldsymbol{k}'},
\label{eq:GF_Kanamori}
\end{align}
where
\begin{equation}
    \hat{h}_{\boldsymbol{k}\boldsymbol{k}'} = \hat{\xi}_{\boldsymbol{k}}\delta_{\boldsymbol{k}\boldsymbol{k}'}\sigma^0 + \hat{\phi}_{\boldsymbol{k}-\boldsymbol{k}'}\sigma^0
-
\hat{\boldsymbol{M}}_{\boldsymbol{k}-\boldsymbol{k}'}
\cdot
\boldsymbol{\sigma}
\label{eq:action_Hamiltonian}
\end{equation}
is the general Hubbard-Stratonovich Hamiltonian.

We now perform the saddle-point (mean-field) approximation of the effective action in Eq.~\eqref{eq:Seff_Kanamori}.
The saddle point is defined by
\begin{align}
\frac{\delta Z}{\delta \phi_{\boldsymbol{q}\mu\eta}} = 0,
\qquad
\frac{\delta Z}{\delta \mathbf{M}_{\boldsymbol{q}\mu\eta}} = 0.
\end{align}
Using Eqs.~\eqref{eq:S_ferm} and \eqref{eq:S_bos}, before integrating out fermions, we obtain
\begin{align}
2\sum_{\alpha\beta}
(U_c^{-1})^{\mu\eta}_{\alpha\beta}
\phi_{\boldsymbol{q}\alpha\beta}
&=
\langle n_{\boldsymbol{q}\eta\mu} \rangle,
\\[6pt]
2\sum_{\alpha\beta}
(U_s^{-1})^{\mu\eta}_{\alpha\beta}
\boldsymbol{M}_{\boldsymbol{q}\alpha\beta}
&=
\langle \boldsymbol{S}_{\boldsymbol{q}\eta\mu} \rangle,
\end{align}
where averages are fermionic.
Thus, the auxiliary fields acquire the physical interpretation
\begin{align}
\phi_{\boldsymbol{q}\mu\eta}
&=
\frac{1}{2}
\sum_{\alpha\beta}
(U_c)^{\mu\eta}_{\alpha\beta}
\langle n_{\boldsymbol{q}\alpha\beta} \rangle,
\\[6pt]
\boldsymbol{M}_{\boldsymbol{q}\mu\eta}
&=
\frac{1}{2}
\sum_{\alpha\beta}
(U_s)^{\mu\eta}_{\alpha\beta}
\langle \boldsymbol{S}_{\boldsymbol{q}\alpha\beta} \rangle.
\end{align}
Notice that $\phi_{0\mu\eta}$ is the known Hartree correction.
$\boldsymbol{M}_{\boldsymbol{q}\mu\eta}$ and $\phi_{\boldsymbol{q}\mu\eta}$ describe orbital-resolved magnetic order and charge density modulations at wavevector $\boldsymbol{q}$, respectively.
We remark that $\boldsymbol{q}$ is not the same for the charge and magnetic order parameters.
Here, the magnetic wave vectors are $\boldsymbol{q}=\boldsymbol{Q}$ and the charge are $\tilde{\boldsymbol{Q}}$ for all sets of $\{\bbs{Q}_i\}$ and $\{\tilde{\bbs{Q}}_i\}$ enclosing an enlarged unit cell.
We also restrict to static, orbital-diagonal saddle-point fields such that
\begin{align}
\phi_{\boldsymbol{q}\mu\eta}(\tau) &\rightarrow \phi_{\mu}\delta(\tau)\delta_{\mu\eta}\delta_{\boldsymbol{q}\tilde{\boldsymbol{Q}}}\\
\boldsymbol{M}_{\boldsymbol{q}\mu\eta}(\tau) &\rightarrow \boldsymbol{M}_{\mu}\delta(\tau)\delta_{\mu\eta}\delta_{\boldsymbol{q}\boldsymbol{Q}}.
\end{align}
We then define the Nambu operators
\begin{align}
\Psi_{\boldsymbol{k}\sigma}
=
\left(
c_{\boldsymbol{k}\mu\sigma},
c_{(\boldsymbol{k}+\boldsymbol{Q})\mu\sigma},\dots,
c_{(\boldsymbol{k}+\tilde{\boldsymbol{Q}})\mu\sigma},\dots
\right)^{T}
\label{eq:spinor_kanamori}
\end{align}
to write the Hamiltonian of Eq.~\eqref{eq:action_Hamiltonian} as the form of Eq.~\eqref{eq:HMF}.
In doing so, we are increasing the number of sites and orbitals in the Hamiltonian, which is the effect of unit cell folding, implying that the Brillouin zone changes.
To avoid double-counting momenta, a reduced Brillouin zone (RBZ) is defined with a fractional area of the normal-state Brillouin zone.

The trace-log term in Eq.~\eqref{eq:Seff_Kanamori} becomes
\begin{align}
\mathrm{Tr}\ln G^{-1}
=
\sum_{\boldsymbol{k}\in RBZ}
\sum_{\nu}
\sum_{\omega_m}
\ln\left(-i\omega_m + \epsilon_{\boldsymbol{k}\nu}\right).
\end{align} 
The Hamiltonian matrix is given in Eq.~\eqref{eq:action_Hamiltonian}, with eigenvalues and eigenvectors defined as $\hat h(\boldsymbol k)|\boldsymbol{k}\nu\rangle=\epsilon_{\boldsymbol k\nu}|\boldsymbol{k}\nu\rangle$ where $\nu$ indexes all the eigenvalues, including spin, which is not a quantum number in non-collinear magnetism or in models with spin-orbit coupling.
We remind that $\epsilon_{\boldsymbol k\nu}$ includes the chemical potential.
Performing the Matsubara summation for the fermionic frequencies $i\omega_m$,
\begin{align}
\sum_{\omega_m}
\ln\left(-i\omega_m + E\right)
=
\ln\left(1+e^{-\beta E}\right).
\end{align}
With this, and noticing that, in the saddle-point solution, Eq.~\eqref{eq:Z} implies the free energy
\begin{equation}
\begin{split}
F
=
&-\sum_{\tilde{\bbs Q} \mu\eta}
\phi_{-\tilde{\bbs Q} \mu}
(U_c^{-1})^{\mu\mu}_{\eta\eta}
\phi_{\tilde{\bbs Q} \eta}
+
\sum_{\bbs Q \mu\eta}
\boldsymbol{M}_{-\bbs Q \mu}\cdot
(U_s^{-1})^{\mu\mu}_{\eta\eta}
\boldsymbol{M}_{\bbs Q \eta}\\
&-
\frac{1}{\beta}
\sum_{\boldsymbol k\in RBZ}
\sum_{\nu}
\ln\left(
1+e^{-\beta \epsilon_{\boldsymbol k\nu}}
\right).\label{eq:FMF_Kanamori}
\end{split}
\end{equation}

We now derive the jacobian of the mean-field free energy obtained in the previous subsection, within the collinear and intraorbital approximation.
The fermionic Hamiltonian is diagonalized in the reduced Brillouin zone (RBZ) using the spinor in Eq.~\eqref{eq:spinor_kanamori}.
In its turn, the mean-field free energy reads as Eq.~\eqref{eq:FMF_Kanamori}.
To derive the jacobian, necessary to efficiently minimize the grand potential, we compute derivatives of $F$ with respect to ${M}_{\bbs Q\mu}^a$ ($a=x,y,z$) and $\phi_{\tilde{\bbs Q}\mu}$.
We use the identity
\begin{align}
\frac{\partial}{\partial \lambda}
\left(
-\frac{1}{\beta}
\sum_{\boldsymbol{k}\nu}
\ln(1+e^{-\beta \epsilon_{\boldsymbol{k}\nu}})
\right)
=
\sum_{\boldsymbol{k}\nu}
f(\epsilon_{\boldsymbol{k}\nu})
\frac{\partial \epsilon_{\boldsymbol{k}\nu}}{\partial \lambda},
\label{eq:free_energy_derivative_identity}
\end{align}
where $f(E)=1/(e^{\beta E}+1)$ is the Fermi function.
Using the Hellmann-Feynman theorem,
\begin{align}
\frac{\partial \epsilon_{\boldsymbol{k}\nu}}{\partial \lambda}
=
\left\langle \boldsymbol{k} \nu
\left|
\frac{\partial \hat{h}(\boldsymbol{k})}{\partial \lambda}
\right|
\boldsymbol{k} \nu
\right\rangle.
\label{eq:HF_theorem}
\end{align}

We compute
\begin{align}
\frac{\partial F}{\partial M_{\bbs Q\mu}^a}.
\end{align}
The bosonic contribution gives
\begin{equation}
\begin{split}
\frac{\partial}{\partial M_{\bbs Q\mu}^a}
\left(
\sum_{\bbs Q'\eta\alpha a}
M_{\bbs Q'\eta}^a
(U_s^{-1})^{\eta\eta}_{\alpha\alpha}
M_{-\bbs Q'\alpha}^a
\right)
=\\
2\sum_{\eta}
(U_s^{-1})^{\mu\mu}_{\eta\eta}
M_{\bbs Q\eta}^a.
\end{split}
\label{eq:bosonic_derivative_M}
\end{equation}
The fermionic contribution yields
\begin{align}
\sum_{\boldsymbol{k} \nu}
f(\epsilon_{\boldsymbol{k}\nu})
\left\langle \boldsymbol{k} \nu 
\left|
\frac{\partial \hat{h}(\boldsymbol{k})}{\partial M_{\bbs Q\mu}^a}
\right|
\boldsymbol{k} \nu 
\right\rangle.
\end{align}
We define
\begin{align}
\hat{\Delta}_{\bbs Q\mu}^a
=
-\frac{\partial \hat{h}(\boldsymbol{k})}{\partial M_{\bbs Q\mu}^a}.
\end{align}
For the single-$\bbs Q_Y$ case, for example,
\begin{align}
    \hat{\Delta}_{\mu}^a=
\begin{pmatrix}
0 & \hat{1}_{\mu} \\
\hat 1_{\mu} & 0
\end{pmatrix}\otimes\sigma^a,
\label{eq:delta}
\end{align}
where $\hat 1_{\mu}$ is 1 in the diagonal orbital matrix element $\mu$ and zero in all the other matrix elements.
The second equality above exemplifies the case of a single ordering vector $\bbs Q$, but the matrix $\hat{\Delta}^a_{\mu}$ can contain multiple ordering vectors.
The full derivative reads
\begin{equation}
\begin{split}
\frac{\partial F}{\partial M_{\bbs Q\mu}^a}
&=
2\sum_{\eta}
(U_s^{-1})^{\mu\mu}_{\eta\eta}
M_{\bbs Q\eta}^a\\
&-
\frac{1}{N}
\sum_{\boldsymbol{k}\in RBZ}
\sum_{\nu}
\langle \boldsymbol{k} \nu |
\hat{\Delta}_{\bbs Q\mu}^a
| \boldsymbol{k} \nu \rangle
f(\epsilon_{\boldsymbol{k}\nu}).
\end{split}
\label{eq:gap_mag}
\end{equation}

Similarly, we compute
\begin{align}
\frac{\partial F}{\partial \phi_{\tilde{\bbs Q}\mu}}.
\end{align}
The bosonic part gives
\begin{equation}
\begin{split}
\frac{\partial}{\partial \phi_{\tilde{\bbs Q}\mu}}
\left(
-\sum_{\tilde{\bbs Q}'\eta\alpha}
\phi_{\tilde{\bbs Q}'\eta}
(U_c^{-1})^{\eta\eta}_{\alpha\alpha}
\phi_{\tilde{\bbs Q}'\alpha}
\right)
=\\
-2\sum_{\eta}
(U_c^{-1})^{\mu\mu}_{\eta\eta}
\phi_{\tilde{\bbs Q}\eta}.
\end{split}
\end{equation}
The fermionic part gives
\begin{align}
\sum_{\boldsymbol{k} \nu}
f(\epsilon_{\boldsymbol{k}\nu})
\left\langle \boldsymbol{k} \nu
\left|
\frac{\partial \hat{h}(\boldsymbol{k})}{\partial \phi_{\mu}}
\right|
\boldsymbol{k} \nu
\right\rangle.
\end{align}
We define
\begin{align}
\hat{N}_{\tilde{\bbs Q}\mu}
=
\frac{\partial \hat{h}(\boldsymbol{k})}{\partial \phi_{\tilde{\bbs Q}\mu}}.
\end{align}
For the single-$\bbs Q_Y$ and $\tilde{\bbs Q}_0=2\bbs Q_Y=0$, for example,
\begin{align}
\hat{N}_{\mu}
=
\begin{pmatrix}
\hat{1}_{\mu} & 0 \\
0 & \hat 1_{\mu}
\end{pmatrix}
\otimes\sigma^0.
\end{align}
Hence,
\begin{equation}
\begin{split}
\frac{\partial F}{\partial \phi_{\tilde{\bbs Q}\mu}}
&=
-2\sum_{\eta}
(U_c^{-1})^{\mu\mu}_{\eta\eta}
\phi_{\tilde{\bbs Q}\eta}\\
&+
\frac{1}{N}
\sum_{\boldsymbol{k}\in RBZ}
\sum_{\nu}
\langle \boldsymbol{k} \nu |
\hat{N}_{\tilde{\bbs Q}\mu}
| \boldsymbol{k} \nu \rangle
f(\epsilon_{\boldsymbol{k}\nu}).
\end{split}
\label{eq:gap_charge}
\end{equation}

The chemical potential $\mu$ must be determined self-consistently in order to fix the total electron density. 
It enters the mean-field Hamiltonian through
\begin{align}
\xi_{\boldsymbol{k}}^{\mu\eta}
=
\epsilon_{\boldsymbol{k}}^{\mu\eta}
-
\mu\delta_{\mu\eta}.
\end{align}
Since the grand potential $\Omega=F+\mu n$, minimization on $\mu$ implies
\begin{align}
n
=
-\,\frac{\partial F}{\partial \mu}.
\end{align}
Using
\begin{align}
\frac{\partial \epsilon_{\boldsymbol{k}\nu}}{\partial \mu}
=
\left\langle \boldsymbol{k} \nu
\left|
\frac{\partial \hat{h}(\boldsymbol{k})}{\partial \mu}
\right|
\boldsymbol{k} \nu
\right\rangle=
-{1},
\end{align}
we obtain
\begin{align}
-\frac{\partial F}{\partial \mu}
=
\frac{1}{N}
\sum_{\boldsymbol{k}\in RBZ}
\sum_{\nu}
f(\epsilon_{\boldsymbol{k}\nu}).
\end{align}
Therefore, the chemical potential equation fixing the filling $n$ is
\begin{align}
n
=
\frac{1}{N}
\sum_{\boldsymbol{k}\in RBZ}
\sum_{\nu}
f(\epsilon_{\boldsymbol{k}\nu}).\label{eq:mu_kanamori}
\end{align}
Eqs.~\eqref{eq:gap_mag} and \eqref{eq:gap_charge} are provided to the minimization algorithm and \eqref{eq:mu_kanamori} is solved implicitly at every iteration of the minimization procedure, self-consistently for all $\boldsymbol{M}_{\bbs Q\mu}$, $\phi_{\tilde{\bbs Q}\mu}$, and $\mu$.

We numerically implement the non-interacting model of Eq.~\eqref{eq:H0_sublattice} using the TRIQS library \cite{ParcolletTRIQSToolboxForResearch2015}.
This code is actively maintained and efficiently deals with real-space hopping parameters and intricate details of all sorts of lattice symmetries.

To solve the mean-field equations, which are obtained by equating the jacobian of the free energy functional to zero that is, $\boldsymbol \nabla F = 0$, we find the extrema of the multivariate grand potential.
We first remark that from Eq.~\eqref{eq:FMF_Kanamori}, the charge channel $\phi_{\tilde{\bbs Q}\mu}$ tends to decrease the free energy, implying that the extrema in $\phi_{\tilde{\bbs Q}\mu}$ are maxima.
In fact, the $\phi_{{\bbs 0}\mu}$ channel corresponds to Hartree shifts, which are always non-zero.
In the square-lattice Hubbard model, for example, $\phi_{{\bbs 0}\mu}=U/2$.
On the other hand, the spin-channel fields $\bbs M_{\bbs Q\mu}$ correspond to minima of the free energy, as a usual Landau order parameter.
We use the Broyden-Fletcher-Goldfarb-Shanno algorithm \cite{Broyden1970DoubleRankMinimization,Fletcher1970VariableMetric,Goldfarb1970VariableMetric,Shanno1970ConditioningQuasiNewton} as implemented by the standard \textit{python} library \textit{scipy} to find minima along $\bbs M_{\bbs Q\mu}$ and maxima along $\phi_{\tilde{\bbs Q}\mu}$.
The function to be extremized is the grand potential $\Omega=F+\mu n$, where the free energy $F$ is given by Eq.~\eqref{eq:FMF_Kanamori}.
As can be noted in Eq.~\eqref{eq:mu_kanamori}, the equation for the chemical potential $\mu$ is a simpler one-dimensional problem.
We, therefore, solve the chemical potential constraint implicitly by the nearly-exact one-dimensional Brent’s method \cite{Brent1973} (\textit{brentq} function of \textit{scipy.optimize}).
In our openly available code, an example of application is given for the $\boldsymbol{Q}=(\pi,\pi)$ square lattice Hubbard model, where the free energy and its jacobian vector field can be visualized in a two-dimensional map and the respective minima and maxima clearly spotted.

\section{Evolution of order parameters with $\Delta_{\text{CO}}/U$}
\label{appendix:co}
\begin{figure}[t]
\begin{center}
\includegraphics[width=1.0\columnwidth]{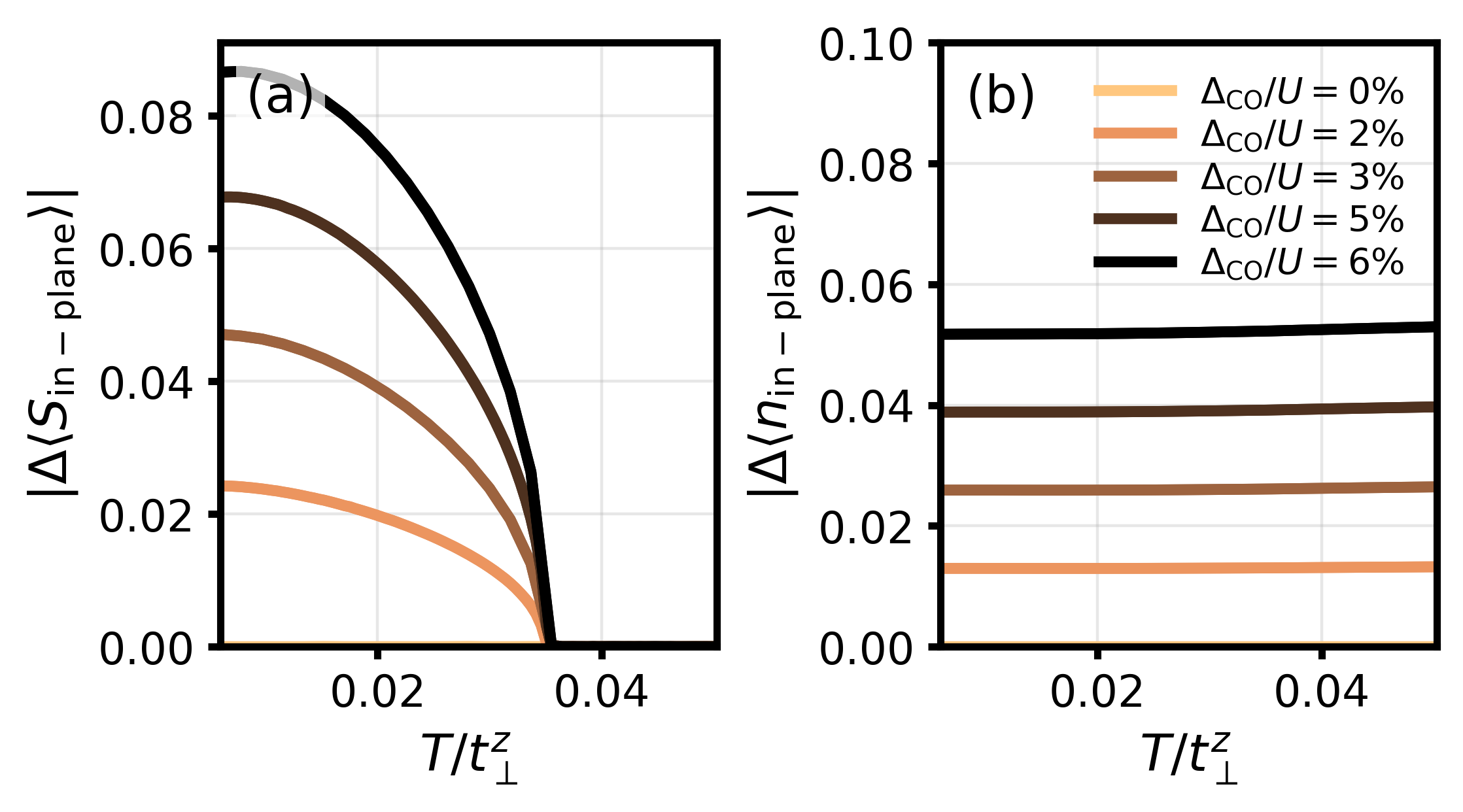}
\caption{ Difference in spin-densities (a) and charge densities (b) of the in-plane Ni sites in the four-Ni unit cell of the orthorhombic bilayer nickelate structure.
Colors denote different $\Delta_{\text{CO}}/U$ ratios.
}
\label{fig:op_co}
\end{center}
\end{figure}

The preceding charge order generated by the $I4/mmm\rightarrow Am2m$ structural transition affects the temperature evolution of the charge and, eventually, magnetic order parameters.
In Fig.~\ref{fig:op_co}, we show the difference of spin and charge densities between inequivalent in-plane Ni atoms.
Without the preceding charge order ($\Delta_{\text{CO}}/U=0$), for the set of parameters used in the simulation, there is no unbalance between spin densities inside the SDW phase.
The dependence on $U$ and $J_H$ of the phase diagram was discussed by the authors in \cite{Braz2026DensityWaves}.
Depending on $U$ and $J_H$, the system can develop a CDW inside the SDW phase.
When $\Delta_{\text{CO}}/U\neq0$, however, the situation is simpler and the only option is developing a SDW phase with unequal spin densities.
Specific tuning of $U$ and $J_H$ becomes unimportant and only qualitatively changes the phase diagram when $U$ is large enough to produce a magnetic insulator.
As shown in Fig.~\ref{fig:op_co}, increasing $\Delta_{\text{CO}}/U$ enhances the spin modulation in the stripe order (a).

\section{Electron-light coupling and the Raman vertex}
\label{appendix:e-l_coupling}
Scattering experiments are based on the absorption and emission of light in and from the material.
Consider the following experiment based on Ref. \cite{Abrikosov1961RamanScattering}.
An incoming photon with momentum $e\boldsymbol{A}/c$ ($e$ electron charge, $c$ light speed and $\boldsymbol{A}$ electromagnetic potential field) hits a material with electronic quadratic Hamiltonian
\begin{equation}
H_{\text{eff}}
=\sum_{\boldsymbol{k}}\sum_{\mu\eta}\sum_{\sigma\sigma'}
c^{\dagger}_{\bbs k\mu\sigma}
h_{\bbs k\bbs k\mu\eta\sigma\sigma'}
c_{\bbs k\eta\sigma'}.
\end{equation}
At first approximation, we could consider that the electron-photon coupling is very quick such that no retardation is relevant and only an instantaneous momentum exchange is experienced by the electrons in the material.
As we will see, this approximation introduces a momentum-dependent effective mass for the band dispersions.
The Hamiltonian after the electronmagnetic interaction takes place becomes
\begin{equation}
H_{\text{eff-l}}
=\sum_{\boldsymbol{k}}\sum_{\mu\eta}\sum_{\sigma\sigma'}
c^{\dagger}_{\bbs k\mu\sigma}
h_{(\bbs k-e\bbs A/c)(\bbs k-e\bbs A/c)\mu\eta\sigma\sigma'}
c_{\bbs k\eta\sigma'}.
\end{equation}
We now expand the matrix $\hat{h}$ around $\bbs k$ and in powers of $\bbs A$ as
\begin{equation}
\begin{split}
H_{\text{eff-l}}
&=\sum_{\boldsymbol{k}}\sum_{\mu\eta}\sum_{\sigma\sigma'}
c^{\dagger}_{\bbs k\mu\sigma}
\big[
h_{\bbs k\bbs k\mu\eta\sigma\sigma'}
-\frac{e}{c}\bbs \nabla h_{\bbs k\bbs k\mu\eta\sigma\sigma'}\cdot\bbs A \\
&+\frac{e^2}{2c^2}\sum_{ij}A_i \frac{\partial^2 h_{\bbs k\bbs k\mu\eta\sigma\sigma'}}{\partial k_i\partial k_j}A_j
+\mathcal{O}(A^3)
\big]
c_{\bbs k\eta\sigma'},
\end{split}
\end{equation}
where the indices $i,j$ run over the spatial directions in a suitable coordinate system.
We readily identify the first term above as the bare system $H_{\text{eff}}$ such that the electron-light interaction reduces to
\begin{equation}
\begin{split}
H_{\text{e-l}}
&=\sum_{\boldsymbol{k}}\sum_{\mu\eta}\sum_{\sigma\sigma'}
c^{\dagger}_{\bbs k\mu\sigma}
\big[
-\frac{e}{c}\bbs \nabla h_{\bbs k\bbs k\mu\eta\sigma\sigma'}\cdot\bbs A \\
&+\frac{e^2}{2c^2}\sum_{ij}A_i \frac{\partial^2 h_{\bbs k\bbs k\mu\eta\sigma\sigma'}}{\partial k_i\partial k_j}A_j
\big]
c_{\bbs k\eta\sigma'}
\end{split}
\end{equation}
%
\begin{figure}[t]
\begin{center}
\includegraphics[width=1.0\columnwidth]{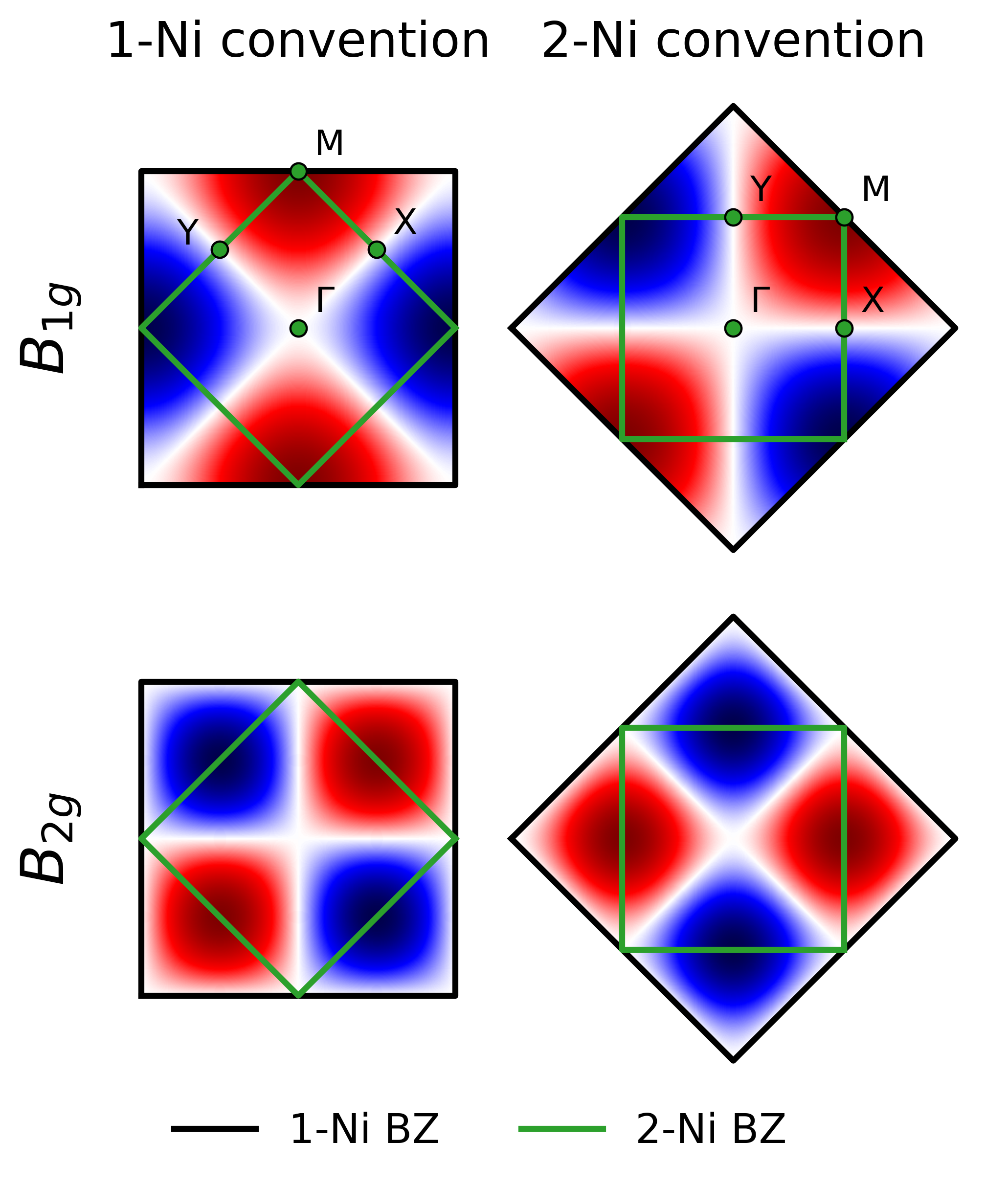}
\caption{Illustration of the effect of Brillouin zone conventions for the interpretation of $B_{1g}$ and $B_{2g}$ symmetry channels.
The 1-Ni convention, used in the experiments, is twisted of 45$^\circ$ in the 2-Ni convention, making the $B_{1g}$ and $B_{2g}$ symmetry channels exchange.
The form factors used in this sketch are artificial.}
\label{fig:bz_conventions}
\end{center}
\end{figure}
up to third order in the incident light electromagnetic field.
The linear term is a velocity and the second-order term plays the role of an effective mass induced by the light-matter interaction.
The present expansion procedure is a good approximation as whenever the characteristic lattice length $L=V^{1/3}$ estimated from the unit cell voluem $V$ is such that $eA/c\ll 1/L$.
Finally, we note that in non-resonant Raman scattering, there must be both ingoing and outgoing photons, which is only realized in the second-order term in $A$.
The first-order term above is, therefore, only relevant in resonant Raman scattering.
In sum, Raman scattering can be approximated by the effective quadratic Hamiltonian
\begin{equation}
H_{\text{e-l}}
=\sum_{\boldsymbol{k}}\sum_{\mu\eta}\sum_{\sigma\sigma'}
\gamma^{\mu\sigma\eta\sigma'}_{\bbs k}
c^{\dagger}_{\bbs k\mu\sigma}
c_{\bbs k\eta\sigma'},
\end{equation}
where
\begin{equation}
\gamma^{\mu\sigma\eta\sigma'}_{\bbs k}
=
\frac{e^2}{2c^2}\sum_{ij}A_i \frac{\partial^2 h_{\bbs k\bbs k\mu\eta\sigma\sigma'}}{\partial k_i\partial k_j}A_j.
\label{eq:raman_vertex}
\end{equation}
Therefore, the light-matter interaction can be interpreted as a mass-renormalization effect, and the Raman scattering matrix plays the role of a hopping matrix.

The potential vector $\bbs A$ is a tunable experimental parameter, depending on the incoming light polarization.
The Raman vertex [Eq.~\eqref{eq:raman_vertex}] at different polarization directions are typically decomposed into symmetry channels of the underlying crystal, but the choice of such basis is arbitrary.
{Here, we follow the choice made by the available experiments and consider the $D_{4h}$ point group, in which the relevant symmetry channels are the $A_{1g}$, coming from $\boldsymbol{A}=(1,0,0)$ plus $(0,1,0)$ contributions, the $B_{2g}$, coming from $(1,0,0)$ minus $(0,1,0)$ contributions, and the $B_{1g}$, coming from $(1,1,0)$ minus $(1,-1,0)$ vertices.
We highlight that the $B_{2g}$ and $B_{1g}$ nomenclatures are chosen to match those used by the available experimental works.}
We remark that the 2-Ni Brillouin zone convention adopted in the present work, as illustrated in Fig. ~\ref{fig:bz_conventions}, has a momentum dependence which exchanges $B_{1g}$ and $B_{2g}$ channels as compared with the 1-Ni convention.

\begin{figure}[t]
\begin{center}
\includegraphics[width=1.0\columnwidth]{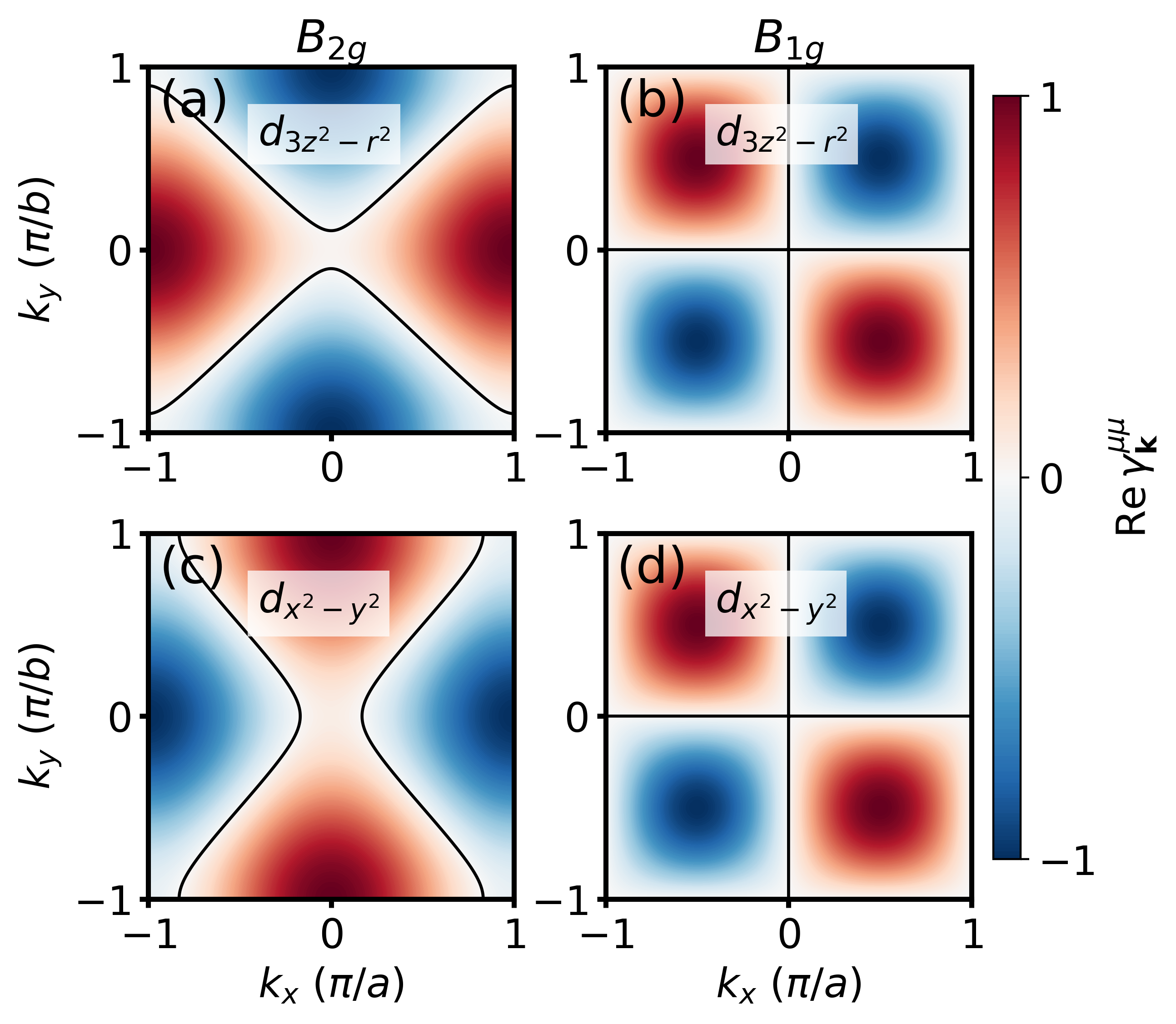}
\caption{ Matrix elements of the Raman vertex for respective orbitals $\mu$ of the orthorhombic-structure two-orbital bilayer nickelate system.
Black lines mark sign-changing Raman vertex contours.}
\label{fig:B1g_B2g}
\end{center}
\end{figure}

In what follows, we will consider the Hamiltonian $h_{\bbs k\bbs k\mu\eta\sigma\sigma'}$ within mean field.
Even inside the ordered state, the mean-field part of the Hamiltonian is momentum independent and, therefore, does not affect the momentum derivatives present in the Raman scattering matrix.
Also, if the non-interacting Hamiltonian is spin-degenerate, so $\gamma^{\mu\sigma\eta\sigma'}_{\bbs k}\rightarrow\gamma^{\mu\eta}_{\bbs k}\delta_{\sigma\sigma'}$.

Fig.~\ref{fig:B1g_B2g} shows the orbital dependence of the Raman vertices for the $B_{2g}$ (left column) and $B_{1g}$ (right column) channels.
We remark that the $B_{2g}$ channel features a different sign structure for each orbital and a consequent different interplay with the broken $C_4$ symmetry of the orthorhombic crystal structure of the bilayer nickelate.
The sensitiveness of the $B_{2g}$ Raman vertex to the orthorhombic distortion has been discussed in the context of a square-lattice model \cite{Branch1995RamanContinuum}.

\section{Raman response function}
\label{appendix:derivation}
Here, we provide a detailed derivation of the Raman response function from the field theory developed in Ref. [our paper].
Based on the fluctuation-dissipation theorem, a response function $\chi^{\gamma\bar{\gamma}}_{\boldsymbol{q}}(\omega)$ relates to the observable structure factor $S^{\gamma\bar{\gamma}}_{\boldsymbol{q}}(\omega)$ as
\begin{equation}
S^{\gamma\bar{\gamma}}_{\boldsymbol{q}}(\omega) = [1+n_B(\omega)]\text{Im} \chi^{\gamma\bar{\gamma}}_{\boldsymbol{q}}(\omega).
\end{equation}
Here, the real frequency $\omega$ is defined from the analytical continuation $i\omega_m\rightarrow\omega+i0^+$, with $\omega_m$ a bosonic Matsubara frequency.
The response function is defined in terms of ``effective charge-density fluctuations'' $\tilde{{\rho}}^{\gamma}_{\boldsymbol{q}}$,
\begin{equation}
\begin{split}
\chi^{\gamma\bar{\gamma}}_{\boldsymbol{q}}(\omega) &= \int_0^{\beta}d\tau e^{-i(\omega+i0^+)\tau} \\
&\times\big[
\langle\tilde{{\rho}}^{\gamma}_{\boldsymbol{q}}(\tau)\tilde{{\rho}}^{\bar{\gamma}}_{-\boldsymbol{q}}(0)\rangle
-
\langle\tilde{{\rho}}^{\gamma}_{\boldsymbol{q}}(\tau)\rangle\langle\tilde{{\rho}}^{\bar{\gamma}}_{-\boldsymbol{q}}(0)\rangle
\big].
\end{split}
\label{eq:raman_def}
\end{equation}
where
\begin{equation}
\tilde{{\rho}}^{\gamma}_{\bbs{q}}(\tau) = \sum_{\bbs k \sigma \mu}\gamma^{\mu\mu}_{\bbs k} c^{\dagger}_{(\bbs k + \bbs q)\mu\sigma}(\tau)c_{\bbs k \mu\sigma}(\tau).
\end{equation}
The function $\gamma^{\mu\mu}_{\bbs k}$ depends on the incident light polarization and the curvature of the Hamiltonian \cite{Devereaux1999ElectronicRaman}.
In what follows, we generalize the approach of Schmalian and Fernandes.

We now move to explicitly compute the Raman response function.
Our strategy is to couple an external source field conjugate to the weighted densities $\tilde{{\rho}}^{\gamma}_{\boldsymbol{q}}$.
First, we note that without sources, the action is given by Eq.~\eqref{eq:action}, to which we add a source term
\begin{equation}
S_{\text{source}}[v^{\gamma}] = \int_0^{\beta}d\tau
\sum_{\boldsymbol{k} \boldsymbol{k}'\mu\eta\sigma\sigma'}
c^{\dagger}_{\boldsymbol{k}\mu\sigma}
V_{v\bbs k\bbs k\mu\eta\sigma\sigma'}^{\gamma}
c_{\boldsymbol{k}'\eta\sigma'},
\end{equation}
\begin{equation}
V_{v\bbs k\bbs k\mu\eta\sigma\sigma'}^{\gamma} = v^{\gamma}_{\bbs k-\bbs k'}\gamma^{\mu\eta}_{\bbs k}\delta_{\sigma\sigma'}\delta_{\mu\eta}.
\end{equation}
Therefore, the partition function associated to $S+S_{\text{source}}$ or equivalently, $S_{\text{eff}}+S_{\text{source}}$ is
\begin{equation}
\begin{split}
Z_{\text{eff+s}}&=\int\mathcal{D}[\hat{\phi},\hat{\boldsymbol{M}}]\mathcal{D}[\hat{c},\hat{c}^\dagger]e^{-S_{\text{eff}}} \\
&\times \exp{\Big(-\int_0^{\beta}d\tau
\sum_{\boldsymbol{k} \boldsymbol{k}'\mu\eta\sigma\sigma'}
c^{\dagger}_{\boldsymbol{k}\mu\sigma}
V_{v\bbs k\bbs k\mu\eta\sigma\sigma'}^{\gamma}
c_{\boldsymbol{k}'\eta\sigma'}\Big)}.
\end{split}
\end{equation}

Next, we prove that the response function is related to the Free energy $F_{\text{eff+s}}=-\beta^{-1}\ln Z_{\text{eff+s}}$ as
\begin{align}
\chi^{\gamma\bar{\gamma}}_{\boldsymbol{q}}(\tau) 
&= -\beta \frac{\delta^2 F_{\text{eff+s}}}{\delta v^{\gamma}_{\bbs q}(\tau)\delta v^{\bar{\gamma}}_{-\bbs q}(0)}\bigg\rvert_{v=0} \\
&=
\frac{\delta}{\delta v_{\bbs q}^{\gamma}(\tau)}
Z_{\text{eff+s}}^{-1}
\frac{\delta Z_{\text{eff+s}}}{\delta v_{-\bbs q}^{\bar{\gamma}}(0)} \\
&=
\frac{1}{Z_{\text{eff+s}}}
\frac{\delta^2 Z_{\text{eff+s}}}{\delta v^{\gamma}_{\bbs q}(\tau)\delta v^{\bar{\gamma}}_{-\bbs q}(0)}\bigg\rvert_{v=0} \nonumber \\
& -
\frac{1}{Z_{\text{eff+s}}^2}
\Big(
\frac{\delta Z_{\text{eff+s}}}{\delta v_{\bbs q}^{{\gamma}}(\tau)}
\Big)
\Big(
\frac{\delta Z_{\text{eff+s}}}{\delta v_{-\bbs q}^{\bar{\gamma}}(0)}
\Big)
\bigg\rvert_{v=0} 
\label{eq:R_derivative}
\end{align}
The second term above cancels all the disconnected diagrams.
Notice that differentiating $Z_{\text{eff+s}}$ with respect to $v^{\gamma}_{\bbs k-\bbs k'}(\tau)$ gives
\begin{equation}
\begin{split}
\frac{1}{Z_{\text{eff+s}}}&
\frac{\delta^2 Z_{\text{eff+s}}}{\delta v^{\gamma}_{\bbs q}(\tau)\delta v^{\bar{\gamma}}_{-\bbs q}(0)}\bigg\rvert_{v=0}=\frac{1}{Z}\int\mathcal{D}[\hat{\phi},\hat{\boldsymbol{M}}]\mathcal{D}[\hat{c},\hat{c}^\dagger]
e^{-S_{\text{eff}}}\\
&\times\Big(-\sum_{\boldsymbol{k}\mu\sigma}
c^{\dagger}_{(\boldsymbol{k}+\bbs q)\mu\sigma}(\tau)
\gamma^{\mu\mu}_{\bbs k}
c_{\boldsymbol{k}\mu\sigma}(\tau) \Big) \\
&\times\Big(-\sum_{\boldsymbol{k}'\eta\sigma'}
c^{\dagger}_{(\boldsymbol{k}'-\bbs q)\eta\sigma'}(0)
\bar{\gamma}^{\eta\eta}_{\bbs k'}
c_{\boldsymbol{k}\mu\sigma}(0) \Big).
\end{split}
\end{equation}
This is exactly the expectation value
\begin{equation}
\begin{split}
&\langle\sum_{\boldsymbol{k}\mu\sigma}
c^{\dagger}_{(\boldsymbol{k}+\bbs q)\mu\sigma}(\tau)
\gamma^{\mu\mu}_{\bbs k}
c_{\boldsymbol{k}\mu\sigma}(\tau)
\sum_{\boldsymbol{k}'\eta\sigma'}
c^{\dagger}_{(\boldsymbol{k}'-\bbs q)\eta\sigma'}(0)
\bar{\gamma}^{\eta\eta}_{\bbs k'}
c_{\boldsymbol{k}\mu\sigma}(0)\rangle \\
&=
\langle\tilde{{\rho}}^{\gamma}_{\boldsymbol{q}}(\tau)\tilde{{\rho}}^{\bar{\gamma}}_{-\boldsymbol{q}}(0)\rangle,
\end{split}
\end{equation}
which relates to the first part of the Raman response function Eq.~\eqref{eq:raman_def}.
The second part is given by the average
\begin{equation}
\begin{split}
&\langle\sum_{\boldsymbol{k}\mu\sigma}
c^{\dagger}_{(\boldsymbol{k}+\bbs q)\mu\sigma}(\tau)
\gamma^{\mu\mu}_{\bbs k}
c_{\boldsymbol{k}\mu\sigma}(\tau)
\rangle\langle
\sum_{\boldsymbol{k}'\eta\sigma'}
c^{\dagger}_{(\boldsymbol{k}'-\bbs q)\eta\sigma'}(0)
\bar{\gamma}^{\eta\eta}_{\bbs k'}
c_{\boldsymbol{k}\mu\sigma}(0)\rangle \\
&=
\langle\tilde{{\rho}}^{\gamma}_{\boldsymbol{q}}(\tau)
\rangle\langle
\tilde{{\rho}}^{\bar{\gamma}}_{-\boldsymbol{q}}(0)\rangle,
\end{split}
\end{equation}
which proves that the response function can be obtained diagrammatically through the sources $v$ within the present action formulation.

Having Eq.~\eqref{eq:R_derivative} in mind, we integrate out fermions by performing a Berezin Gaussian integral resulting in the effective action
\begin{align}
S_{\text{eff+s}}[v^{\gamma}_{\bbs q}] &= 
\frac{1}{2N}
\int_0^\beta d\tau
\sum_{\bbs q}
\Big[
\hat{\phi}_{-\boldsymbol{q}} \hat{U}_c^{-1} \hat{\phi}_{\boldsymbol{q}}
+
\hat{\boldsymbol{M}}_{-\boldsymbol{q}}\cdot \hat{U}_s^{-1} \hat{\boldsymbol{M}}_{\boldsymbol{q}}
\Big] \nonumber\\
&- \Tr\ln\Big[ \hat{G}^{-1} - \hat{V} \Big] \\
&=\frac{1}{2N}
\int_0^\beta d\tau
\sum_{\bbs q}
\Big[
\hat{\phi}_{-\boldsymbol{q}} \hat{U}_c^{-1} \hat{\phi}_{\boldsymbol{q}} \nonumber+
\hat{\boldsymbol{M}}_{-\boldsymbol{q}}\cdot \hat{U}_s^{-1} \hat{\boldsymbol{M}}_{\boldsymbol{q}}
\Big] \nonumber\\
&- \Tr\ln\Big[ \hat{G}^{-1}\Big] \nonumber\\
&+ \sum_{n=1}^{\infty}\frac{\Tr\Big[\big(\hat{G}\hat{V}\big)^n\Big]}{n}\\
&= S_{\text{eff}} 
+ \sum_{n=1}^{\infty}\frac{\Tr\Big[\big(\hat{G}\hat{V}\big)^n\Big]}{n},
\end{align}
where we have used a power-law expansion for the logarithm term.
With the partition function
\begin{equation}
Z_{\text{eff+s}} = 
\int\mathcal{D}[\hat{\phi},\hat{\boldsymbol{M}}]\mathcal{D}[\hat{c},\hat{c}^\dagger]
e^{-S_{\text{eff}}}
\exp{\Big( -\sum_{n=1}^{\infty}\frac{\Tr\Big[\big(\hat{G}\hat{V}\big)^n\Big]}{n} \Big)},
\end{equation}
and using Eq.~\eqref{eq:R_derivative}, the Raman response function becomes
\begin{align}
&\langle\tilde{{\rho}}^{\gamma}_{\boldsymbol{q}}(\tau)
\tilde{{\rho}}^{\bar{\gamma}}_{-\boldsymbol{q}}(0)\rangle
= 
\frac{1}{Z}\int\mathcal{D}[\hat{\phi},\hat{\boldsymbol{M}}]e^{-S_{\text{eff}}}\nonumber\\
&\times
\frac{\delta^2}{\delta v^{\gamma}_{\bbs q}(\tau)\delta v^{\bar{\gamma}}_{-\bbs q}(0)}
\exp{\Big( -\sum_{n=1}^{\infty}\frac{\Tr\Big[\big(\hat{G}\hat{V}\big)^n\Big]}{n} \Big)}
\bigg\rvert_{v=0} \\
&=
\frac{1}{Z}\int\mathcal{D}[\hat{\phi},\hat{\boldsymbol{M}}]\mathcal{D}[\hat{c},\hat{c}^\dagger]
e^{-S_{\text{eff}}}\nonumber\\
&\times\frac{\delta}{\delta v^{\gamma}_{\bbs q}(\tau)}
\exp{\Big( -\sum_{n=1}^{\infty}\frac{\Tr\Big[\big(\hat{G}\hat{V}\big)^n\Big]}{n} \Big)} \nonumber \\
&\times\Big( -\sum_{n=1}^{\infty}{\Tr\Big[\big(\hat{G}\hat{V}\big)^{(n-1)}\hat{G}
\frac{\delta\hat{V}}{\delta v^{\bar{\gamma}}_{-\bbs q}}\Big]} \Big)
\bigg\rvert_{v=0}. \label{eq:first_der}
\end{align}
From the above expression, one can see that the $v=0$ evaluation will suppress all terms of order $n>2$.
Thus,
\begin{equation}
\begin{split}
&\langle\tilde{{\rho}}^{\gamma}_{\boldsymbol{q}}(\tau)
\tilde{{\rho}}^{\bar{\gamma}}_{-\boldsymbol{q}}(0)\rangle = 
\frac{1}{Z}\int\mathcal{D}[\hat{\phi},\hat{\boldsymbol{M}}]
e^{-S_{\text{eff}}}\\
&\times\frac{\delta^2}{\delta v^{\gamma}_{\bbs q}(\tau)\delta v^{\bar{\gamma}}_{-\bbs q}(0)}
\exp{\Big( -\Tr\big(\hat{G}\hat{V}\big) 
-\frac{\Tr\big[\big(\hat{G}\hat{V}\big)^2\big]}{2} \Big)}
\bigg\rvert_{v=0}.
\end{split}
\end{equation}
Defining $\hat{\Gamma}^{\gamma}_{\bbs q}=\frac{\delta\hat{V}}{\delta v^{{\gamma}}_{\bbs q}}$, the perturbation series of $\chi^{\gamma\bar{\gamma}}_{\boldsymbol{q}}(\tau)$ is given in terms of
\begin{equation}
\begin{split}
&\frac{\delta^2\exp{\Big( -\Tr\big(\hat{G}\hat{V}\big) 
-\frac{\Tr\big[\big(\hat{G}\hat{V}\big)^2\big]}{2} \Big)}}{\delta v^{\gamma}_{\bbs q}(\tau)\delta v^{\bar{\gamma}}_{-\bbs q}(0)}
\bigg\rvert_{v=0}\\
&=
\Tr\big(\hat{G}\hat{\Gamma}^{\gamma}_{\bbs q}\big)
\Tr\big(\hat{G}\hat{\Gamma}^{\bar{\gamma}}_{-\bbs q}\big)
-
\Tr\big(\hat{G}\hat{\Gamma}^{\gamma}_{\bbs q}\hat{G}\hat{\Gamma}^{\bar{\gamma}}_{-\bbs q}\big),
\end{split}
\end{equation}
where we have used that the source field is linear in $v$ such that $=\frac{\delta\hat{\Gamma}^{\gamma}_{\bbs q}}{\delta v^{{\gamma}}_{\bbs q}}=0$.
Lastly, we notice that, from Eq.~\eqref{eq:first_der},
\begin{equation}
\frac{\delta\exp{\Big( -\Tr\big(\hat{G}\hat{V}\big) 
-\frac{\Tr\big[\big(\hat{G}\hat{V}\big)^2\big]}{2} \Big)}}{\delta v^{\gamma}_{\bbs q}(\tau)}
\bigg\rvert_{v=0}
=
-\Tr\big(\hat{G}\hat{\Gamma}^{\gamma}_{\bbs q}\big)
\end{equation}
and analogously for the $v_{-\bbs q}^{\bar{\gamma}}(0)$ component, giving the contribution $\langle\tilde{{\rho}}^{\gamma}_{\boldsymbol{q}}(\tau)
\rangle\langle
\tilde{{\rho}}^{\bar{\gamma}}_{-\boldsymbol{q}}(0)\rangle$.
Therefore, the exact expression for the Raman response function is obtained by reassembling these terms to Eq.~\eqref{eq:R_derivative}, that is
\begin{equation}
\chi^{\gamma\bar{\gamma}}_{\boldsymbol{q}}(\tau) =
-\frac{1}{Z}\int\mathcal{D}[\hat{\phi},\hat{\boldsymbol{M}}]
e^{-S_{\text{eff}}}
\Tr\big(\hat{G}\hat{\Gamma}^{\gamma}_{\bbs q}\hat{G}\hat{\Gamma}^{\bar{\gamma}}_{-\bbs q}\big).
\end{equation}

The convenience of the saddle-point solution is that the path integrals over the fields are dropped, because the fields become constants with Dirac deltas setting their exact value.
Thus, within mean-field,
\begin{equation}
\chi^{\gamma\bar{\gamma}}_{\boldsymbol{q}}(\tau) =
-
\Tr\big(\hat{G}\hat{\Gamma}^{\gamma}_{\bbs q}\hat{G}\hat{\Gamma}^{\bar{\gamma}}_{-\bbs q}\big),
\label{eq:response_mf}
\end{equation}
because the denominator $Z_{\text{MF}}=e^{-S_{\text{eff}}}$ cancels with the numerator $e^{-S_{\text{eff}}}$.
This is the main result of this development: the full mean-field response function, with all infinite series of response-function diagrams, given the mean-field solution, is composed dressed bubble and ladder diagrams, where the propagator is the full, self-consistent mean-field one.
We note that even at the normal state, the propagator contains a self-consistently determined self-energy (also called Hartree corrections).
At this point, the usual approximation is to expand the mean-field propagator into a series of bare times self-energy terms.
From this, one considers vertex corrections and all other relevant diagrams.
Here, however, we have direct access to the full propagator and do not need to evoque such approximations.

In what folows, we approach the lengthy task of dealing with indices explicitly.
We first note that
\begin{align}
\hat{\Gamma}^{\gamma}_{\bbs q}(\tau_1-\tau_2)&=\frac{\delta\hat{V}}{\delta v^{{\gamma}}_{\bbs q}(\tau)}\nonumber\\
&=\delta_{(\bbs k - \bbs k')=\bbs q}\gamma^{\mu\eta}_{\bbs k}\delta_{\sigma\sigma'}\delta_{\mu\eta}\delta((\tau_1-\tau_2)-\tau), \\
\hat{\Gamma}^{\gamma}_{-\bbs q}(\tau_1-\tau_2)&=\frac{\delta\hat{V}}{\delta v^{{\gamma}}_{-\bbs q}(0)}\nonumber\\
&=\delta_{(\bbs k - \bbs k')=-\bbs q}\gamma^{\mu\eta}_{\bbs k}\delta_{\sigma\sigma'}\delta_{\mu\eta}\delta((\tau_1-\tau_2)),
\end{align}
so
\begin{align}
&\Tr\Big( \hat{G}\hat{\Gamma}^{\gamma}_{\bbs q}\hat{G}\hat{\Gamma}^{\bar{\gamma}}_{-\bbs q} \Big)\\
&= 
\sum_{\bbs k \bbs k' \bar{\bbs k} \bar{\bbs k}'}\sum_{\mu\eta\bar{\mu}\bar{\eta}}\sum_{\sigma\sigma'\bar{\sigma}\bar{\sigma}'}
G_{\bbs k\bbs k'\mu\eta\sigma\sigma'}(\tau)
\gamma^{\eta\bar{\mu}}_{\bbs k'}
\delta_{(\bbs k' - \bar{\bbs k})=\bbs q} \nonumber\\
&\times\delta_{\sigma'\bar{\sigma}}\delta_{\eta\bar{\mu}}
G_{\bar{\bbs k}\bar{\bbs k}'\bar{\mu}\bar{\eta}\bar{\sigma}\bar{\sigma}'}(-\tau)
\bar{\gamma}^{\bar{\eta}\mu}_{\bar{\bbs k'}}
\delta_{(\bar{\bbs k}' - \bbs k)=-\bbs q}\delta_{\bar{\sigma}'\sigma}\delta_{\bar{\eta}\mu} \\
& = 
\sum_{\bbs k \bbs k'}\sum_{\mu\eta}\sum_{\sigma\sigma'}
G_{\bbs k\bbs k'\mu\eta\sigma\sigma'}(\tau)
\gamma^{\eta\eta}_{\bbs k'}
G_{(\bbs k'+\bbs q)(\bbs k-\bbs q)\eta\mu{\sigma'}{\sigma}}(-\tau)
\bar{\gamma}^{\mu\mu}_{\bbs k-\bbs q}.
\end{align}

If $\hat{G}_{\bbs k \bbs k'}$ is diagonal in momentum, then $\hat{G}_{\bbs k \bbs k'}\rightarrow\hat{G}_{(\bbs k+\{{\bbs Q}_{i}\}) \bbs k'}\delta_{\bbs k'-\bbs k=\{{\bbs Q}_{i}\}}$ imposes
\begin{equation}
\begin{split}
&\Tr\Big( \hat{G}\hat{\Gamma}^{\gamma}_{\bbs q}\hat{G}\hat{\Gamma}^{\bar{\gamma}}_{-\bbs q} \Big) \\
&= \sum_{\{{\bbs Q}_i\}}
\sum_{\bbs k}\sum_{\mu\eta}\sum_{\sigma\sigma'}
G_{\bbs k(\bbs k+\{{\bbs Q}_{i}\})\mu\eta\sigma\sigma'}(\tau)
\gamma^{\eta\eta}_{\bbs k+\{{\bbs Q}_{i}\}} \\
&\times G_{(\bbs k+\{{\bbs Q}_{i}\}+\bbs q)(\bbs k-\bbs q)\eta\mu{\sigma'}{\sigma}}(-\tau)
\bar{\gamma}^{\mu\mu}_{\bbs k-\bbs q}.
\end{split}
\label{eq:trgg}
\end{equation}
In this case, the momentum-space subspace contraction follows a dot-product structure between the first row and first column of the Green's function matrix.
Also, orbital and spin indices contract as a matrix multiplication.

Next, we aim to Fourier transform imaginary time to Matsubara frequency and to take the $\bbs q\rightarrow0$ limit.
First, we note that the Green's functions are
\begin{equation}
G_{\bbs k(\bbs k+\{{\bbs Q}_{i}\})\mu\eta\sigma\sigma'}(\tau)
= \frac{1}{\beta}
\sum_{i\omega_n}e^{-i\omega_n\tau}
G_{\bbs k(\bbs k+\{{\bbs Q}_{i}\})\mu\eta\sigma\sigma'}(i\omega_n),
\end{equation}
where
\begin{equation}
G_{\bbs k(\bbs k+\{{\bbs Q}_{i}\})\mu\eta\sigma\sigma'}(i\omega_n)
=
\sum_\nu \frac{\psi^{\mu\sigma}_{\bbs k\nu}\psi^{\eta\sigma'*}_{(\bbs k+\{\bbs Q_i\})\nu}}{i\omega_n - \epsilon_{\bbs k\nu}},
\end{equation}
where $\psi^{\eta\sigma'*}_{(\bbs k+\{\bbs Q_i\})\nu}$ are eigenvectors of the mean-field Hamiltonian corresponding to the eigenvalue $\epsilon_{\bbs k\nu}$.
It will also be useful to remind that Matsubara summations can be evaluated as contour integrals
\begin{equation}
\frac{1}{\beta}\sum_{i\omega_n}F(i\omega_n) = \oint_C F(z)f(z)dz.
\end{equation}
For $\bbs q\rightarrow0$ and in terms of bosonic Matsubara frequencies $i\omega_m$, Eq.~\eqref{eq:trgg} becomes
\begin{equation}
\begin{split}
&\Tr\Big( \hat{G}\hat{\Gamma}^{\gamma}_{0}\hat{G}\hat{\Gamma}^{\bar{\gamma}}_{0} \Big) = 
\sum_{\{{\bbs Q}_i\}}\sum_{\bbs k}\sum_{\mu\eta}\sum_{\sigma\sigma'}
\gamma^{\eta\eta}_{\bbs k+\{{\bbs Q}_{i}\}}
\bar{\gamma}^{\mu\mu}_{\bbs k}
\frac{1}{\beta^2}\sum_{i\omega_n',i\omega_n} \\
&\times\int_0^\beta e^{-i[\omega_n'-(\omega_m+\omega_n)]\tau}d\tau \\
&\times G_{\bbs k(\bbs k+\{{\bbs Q}_{i}\})\mu\eta\sigma\sigma'}(i\omega_n')
G_{(\bbs k+\{{\bbs Q}_{i}\})\bbs k\eta\mu{\sigma'}{\sigma}}(i\omega_n).
\end{split}
\end{equation}
The frequency part is
\begin{equation}
\begin{split}
&\frac{1}{\beta^2}\sum_{i\omega_n',i\omega_n}
\int_0^\beta e^{-i[\omega_n'-(\omega_m+\omega_n)]\tau}d\tau \\
&\times G_{\bbs k(\bbs k+\{{\bbs Q}_{i}\})\mu\eta\sigma\sigma'}(i\omega_n')
G_{(\bbs k+\{{\bbs Q}_{i}\})\bbs k\eta\mu{\sigma'}{\sigma}}(i\omega_n)
\\
&=\frac{1}{\beta}
\sum_{i\omega_n}
G_{\bbs k(\bbs k+\{{\bbs Q}_{i}\})\mu\eta\sigma\sigma'}(i\omega_n+i\omega_m)\\
&\times G_{(\bbs k+\{{\bbs Q}_{i}\})\bbs k\eta\mu{\sigma'}{\sigma}}(i\omega_n),
\end{split}
\end{equation}
where we have used that both $i\omega_n'=i\omega_m+i\omega_n$ and $i\omega_n'=\epsilon_{\bbs k\nu}$ produce poles.
The canonical bubble-diagram result emerged.
The last Matsubara summation can be expanded as
\begin{equation}
\begin{split}
\frac{1}{\beta}
&\sum_{i\omega_n}
G_{\bbs k(\bbs k+\{{\bbs Q}_{i}\})\mu\eta\sigma\sigma'}(i\omega_n+i\omega_m)
G_{(\bbs k+\{{\bbs Q}_{i}\})\bbs k\eta\mu{\sigma'}{\sigma}}(i\omega_n)
= \\
\frac{1}{\beta}&\sum_{i\omega_n}
\sum_{\nu\nu'}
\frac{\psi^{\mu\sigma}_{\bbs k\nu}\psi^{\eta\sigma'*}_{(\bbs k+\{\bbs Q_i\})\nu}}{i\omega_n - (\epsilon_{\bbs k\nu}-i\omega_m)}
\frac{\psi^{\eta\sigma'}_{(\bbs k+\{\bbs Q_i\})\nu'}\psi^{\mu\sigma*}_{\bbs k\nu'}}{i\omega_n - \epsilon_{\bbs k\nu'}}.
\end{split}
\end{equation}
The case $\nu=\nu',i\omega_m=0$ has to be discriminated from the above Matsubara summation such that
\begin{equation}
\begin{split}
\frac{1}{\beta}&\sum_{i\omega_n}
\sum_{\nu\nu'}
\frac{\psi^{\mu\sigma}_{\bbs k\nu}\psi^{\eta\sigma'*}_{(\bbs k+\{\bbs Q_i\})\nu}}{i\omega_n - (\epsilon_{\bbs k\nu}-i\omega_m)}
\frac{\psi^{\eta\sigma'}_{(\bbs k+\{\bbs Q_i\})\nu'}\psi^{\mu\sigma*}_{\bbs k\nu'}}{i\omega_n - \epsilon_{\bbs k\nu'}} \\
&=
\sum_{\nu\nu'}(1-\delta_{i\omega_m,0}\delta_{\nu,\nu'})
\psi^{\mu\sigma}_{\bbs k\nu}\psi^{\eta\sigma'*}_{(\bbs k+\{\bbs Q_i\})\nu}
\psi^{\eta\sigma'}_{(\bbs k+\{\bbs Q_i\})\nu'}\psi^{\mu\sigma*}_{\bbs k\nu'}\\
&\times\frac{f(\epsilon_{\bbs k\nu})-f(\epsilon_{\bbs k\nu'})}{\epsilon_{\bbs k\nu}-\epsilon_{\bbs k\nu'}-i\omega_m} \\
&+
\delta_{i\omega_m,0}\sum_{\nu}
|\psi^{\mu\sigma}_{\bbs k\nu}|^2
|\psi^{\eta\sigma'*}_{(\bbs k+\{\bbs Q_i\})\nu}|^2\\
&\times\beta 
f(\epsilon_{\bbs k\nu})[f(\epsilon_{\bbs k\nu})-1].
\end{split}
\label{eq:Tr0Tr0}
\end{equation}
Conveniently, the second term in Eq.~\eqref{eq:Tr0Tr0} is purely real, so the imaginary part of this term is
\begin{equation}
\begin{split}
&\text{Im}\Tr\Big( \hat{G}\hat{\Gamma}^{\gamma}_{0}\hat{G}\hat{\Gamma}^{\bar{\gamma}}_{0} \Big)= \\
&\text{Im}\sum_{\{{\bbs Q}_i\}}\sum_{\bbs k}\sum_{\mu\eta}\sum_{\sigma\sigma'}
\gamma^{\eta\eta}_{\bbs k+\{{\bbs Q}_{i}\}}
\bar{\gamma}^{\mu\mu}_{\bbs k} \\
&\times\sum_{\nu\neq\nu'}
\psi^{\mu\sigma}_{\bbs k\nu}\psi^{\eta\sigma'*}_{(\bbs k+\{\bbs Q_i\})\nu}
\psi^{\eta\sigma'}_{(\bbs k+\{\bbs Q_i\})\nu'}\psi^{\mu\sigma*}_{\bbs k\nu'}\\
&\times\frac{f(\epsilon_{\bbs k\nu})-f(\epsilon_{\bbs k\nu'})}{\epsilon_{\bbs k\nu}-\epsilon_{\bbs k\nu'}-i\omega_m}.
\end{split}
\label{eq:Tr0Tr0}
\end{equation}
Finally, we impose analytical continuation $i\omega_m\rightarrow\omega+i0^+$.
In this work, we used $0^+/t_\perp^z\approx .006$ meV.

\section{Raman response in the $A_{1g}$ channel}
\label{appendix:A1g}
\begin{figure}[t]
\begin{center}
\includegraphics[width=1.0\columnwidth]{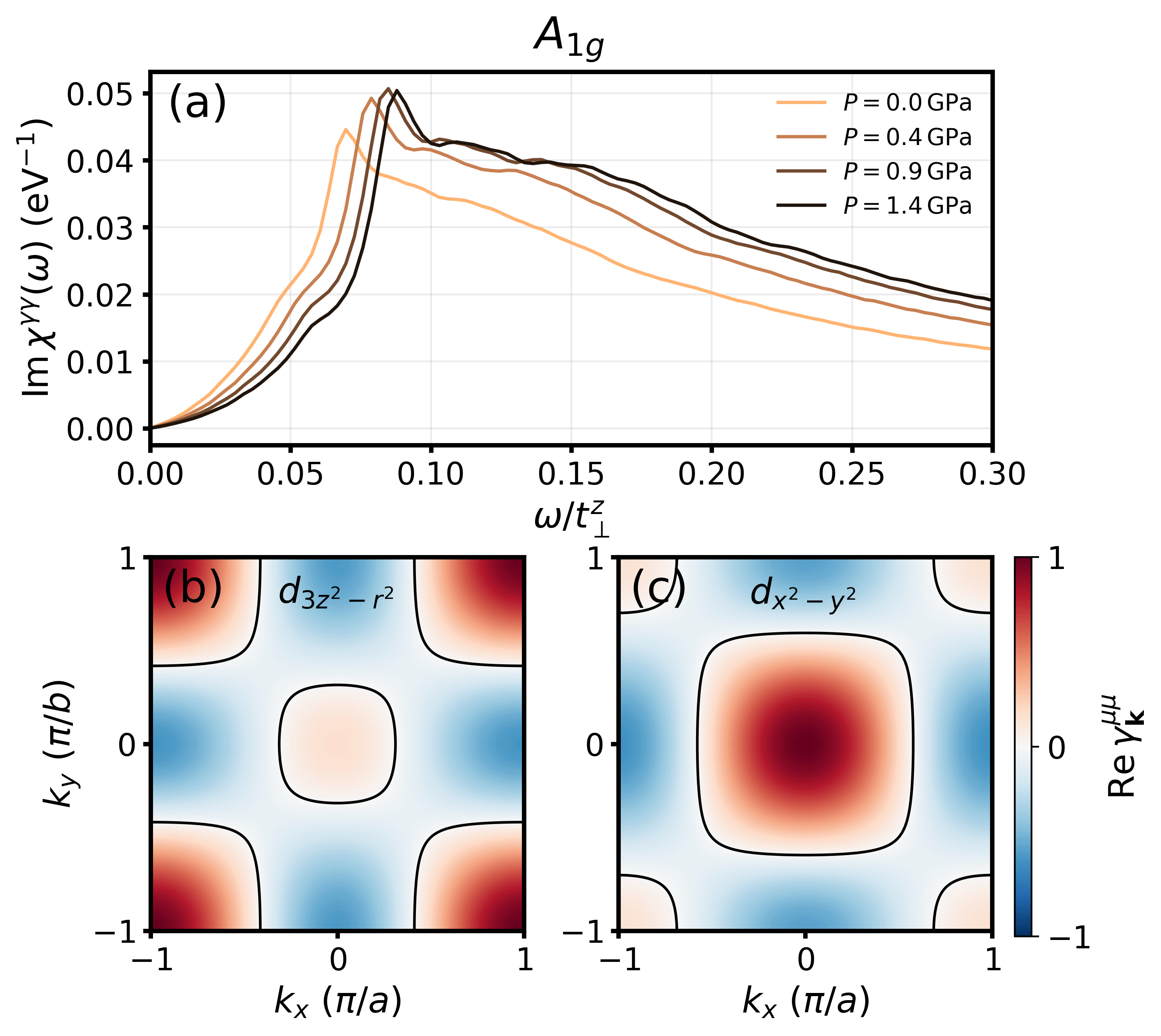}
\caption{ Pressure evolution of the Raman response function (a), and $\mu=d_{3z^2-r^2}$ (b) and $d_{x^2-y^2}$ (c) Raman vertices in the $A_{1g}$ channel. 
$T/T_{\text{SDW}}=62\%$ and $\Delta_{\text{CO}}/U=4\%$ were used.
In (b) and (c), black lines mark sign-changing Raman vertex contours.
}
\label{fig:A1g}
\end{center}
\end{figure}

In a metal, the $A_{1g}$ channel is especially affected by Coulomb screening, which requires proper account of a long-range tail for electron-electron interactions \cite{Dierker1983ElectronicRaman,Devereaux2007InelasticLightScattering}.
In this work, we do not approach this peculiarity and, instead, focus on the $B_{1g}$ and $B_{2g}$ symmetry channels.
We show the unscreened $A_{1g}$ channel Raman response and the Raman vertex in panels (a) and (b-c) of Fig.~\ref{fig:A1g}, respectively.
The $A_{1g}$ response resembles the single-orbital-like $B_{2g}$ response, in close similarity to the single-orbital case explored in the context of cuprate systems \cite{Branch1995RamanContinuum,Devereaux2007InelasticLightScattering}.
On the other hand, at lower pressures, $A_{1g}$ also shows the composite multiorbital peak of $B_{1g}$.
Pressure suppresses the $A_{1g}$ response and especially the composite multiorbital peak at higher energies.

\bibliography{references}

\end{document}